\documentclass[12pt]{article}
\usepackage{amsmath,amssymb,amsfonts,color,graphicx,cite,soul}
\usepackage{colordvi}
\usepackage{subfigure}
\input paperdef
 
\graphicspath{{plots/}}

\newcommand{\MHexp}{125.6}

\evensidemargin \oddsidemargin
\allowdisplaybreaks

\begin{document}
\thispagestyle{empty}

\def\thefootnote{\fnsymbol{footnote}}

\begin{flushright}
MPP--2014--160 \\
arXiv:1404.7074 [hep-ph]
\end{flushright}

\vspace{1cm}

\begin{center}

{\Large\sc {\bf Momentum-dependent two-loop QCD corrections\\[.5em]
 to the neutral Higgs-boson masses in the MSSM}}

\vspace{1cm}

{\sc
S.~Borowka$^{1}$%
\footnote{email: sborowka@mpp.mpg.de}%
, T.~Hahn$^{1}$%
\footnote{email: hahn@mpp.mpg.de}%
, S.~Heinemeyer$^{2}$%
\footnote{email: Sven.Heinemeyer@cern.ch}%
, G.~Heinrich$^{1}$%
\footnote{email: gudrun@mpp.mpg.de}%
~and W.~Hollik$^{1}$%
\footnote{email: hollik@mpp.mpg.de}
}

\vspace*{.7cm}

{\sl
$^1$Max-Planck-Institut f\"ur Physik (Werner-Heisenberg-Institut),\\
F\"ohringer Ring 6, D--80805 M\"unchen, Germany

\vspace*{0.1cm}

$^2$Instituto de F\'isica de Cantabria (CSIC-UC), Santander, Spain
}

\end{center}

\vspace*{0.1cm}

\begin{abstract}
\noindent
Results are presented for the momentum dependent two-loop contributions of
\order{\alt\als} to the masses and mixing effects in the Higgs sector
of the MSSM. They are obtained in the
Feynman-diagrammatic approach using a mixed
on-shell/\DRbar\ renormalization that can directly be matched onto the
higher-order corrections included in the code \fh. 
The new two-loop diagrams are evaluated with the program \sd. 
The combination of the new momentum dependent two-loop contribution with
the existing one- and two-loop corrections in the
on-shell/\DRbar\ scheme leads to an improved prediction of the light
MSSM Higgs boson mass and a correspondingly reduced theoretical
uncertainty. 
We find that the corresponding shifts in the lightest Higgs-boson mass 
$\Mh$ are below~$1 \gev$ in all scenarios considered, but can extend up
 to the level of the current experimental uncertainty.
The results are included in the code \fh.
\end{abstract}

\def\thefootnote{\arabic{footnote}}
\setcounter{page}{0}
\setcounter{footnote}{0}

\newpage


\section{Introduction}

The ATLAS and CMS experiments at CERN have recently discovered a
new boson with a mass around 
$\MHexp \gev$~\cite{ATLASdiscovery,CMSdiscovery}. 
Within the present experimental uncertainties
this new boson behaves like the 
Higgs boson of the Standard Model (SM)~\cite{Moriond14}. However,
the newly discovered 
particle can also be interpreted as the Higgs boson of extended models.
The Higgs sector of the Minimal Supersymmetric
Standard Model (MSSM)~\cite{mssm} with two scalar doublets
accommodates five physical Higgs bosons. In
lowest order these are the light and heavy $\cp$-even $h$
and $H$, the $\cp$-odd $A$, and the charged Higgs bosons $H^\pm$.
The measured mass value, having already reached the level of a precision
observable with an experimental accuracy of about $500 \mev$, plays
an important role in this context. In the MSSM the mass of the light
$\cp$-even Higgs boson, $\Mh$, can directly be predicted from
the other parameters of the model. The accuracy of this prediction 
should at least match the one of the experimental result.

The Higgs sector of the MSSM can be expressed at lowest
order in terms of the gauge couplings, the mass of the $\cp$-odd Higgs boson,
$\MA$, and $\tb \equiv v_2/v_1$,  
the ratio of the two vacuum expectation values. All other masses and
mixing angles can therefore be predicted.
Higher-order contributions can give 
large corrections to the tree-level relations~~\cite{MHreviews,PomssmRep}. 
An upper bound for the
mass of the lightest MSSM Higgs boson of $\Mh \lsim 135 \gev$  was
obtained~\cite{mhiggsAEC}, and the remaining theoretical uncertainty in the
calculation of $\Mh$, from unknown higher-order corrections, was estimated 
to be up to $3 \gev$, depending on the parameter region.
Recent improvements have lead to a somewhat smaller estimate of 
up to $\sim 2 \gev$~\cite{Mh-logresum,ehowp} (see below).

Experimental searches for the neutral MSSM Higgs bosons have been
performed at LEP~\cite{LEPHiggsSM,LEPHiggsMSSM}, placing important
restrictions on the parameter space. At Run~II of the Tevatron the
search was continued, but is now superseeded by the LHC Higgs
searches.
Besides the discovery of a SM Higgs-like boson the LHC searches place
stringent bounds, in particular in the regions of small $\MA$ and large
$\tb$~\cite{CMSHiggsMSSM}. 
At a future linear collider (ILC) a precise determination of the
Higgs boson properties 
(either of the light Higgs boson at $\sim \MHexp \gev$ or heavier MSSM
Higgs bosons within the kinematic reach) will be
possible~\cite{Snowmass13HiggsWP}. 
In particular a mass measurement of the light Higgs boson with an
accuracy below $\sim 0.05 \gev$ is anticipated~\cite{dbd}.
The interplay of the LHC and the ILC in the neutral MSSM Higgs sector 
has been discussed in \citeres{lhcilc,eili}.

For the MSSM%
\footnote{We concentrate here on the case with real parameters. For
the case of complex parameters 
see~\citeres{mhcMSSMlong,mhcMSSM2L,Demir,mhiggsCPXRG1} 
and references therein.}
the status of higher-order corrections to the masses and mixing angles
in the neutral Higgs sector is quite advanced. The complete one-loop
result within the MSSM is known~\cite{ERZ,mhiggsf1lA,mhiggsf1lB,mhiggsf1lC}.
The by far dominant one-loop contribution is the \order{\alt} term due
to top and stop loops ($\alt \equiv h_t^2 / (4 \pi)$, $h_t$ being the
top-quark Yukawa coupling). The computation of the two-loop corrections
has meanwhile reached a stage where all the presumably dominant
contributions are 
available~\cite{mhiggsletter,mhiggslong,mhiggslle,mhiggsFD2,bse,mhiggsEP0,mhiggsEP1,mhiggsEP1b,mhiggsEP2,mhiggsEP3,mhiggsEP3b,mhiggsEP4,mhiggsEP4b,mhiggsRG1,mhiggsRG1a}.
In particular, the \order{\alt\als} contributions to the self-energies -- evaluated in the
Feynman-diagrammatic (FD) as well as in the effective potential (EP)
method -- as well as the \order{\alt^2}, \order{\alb\als}, 
\order{\alt\alb} and \order{\alb^2} contributions  -- evaluated in the EP
approach -- 
are known for vanishing external momenta.  
An evaluation of the momentum dependence at the two-loop level in a pure
\DRbar\ calculation was presented in \citere{mhiggs2lp2}.
A (nearly) full two-loop EP calculation,  
including even the leading three-loop corrections, has also been
published~\cite{mhiggsEP5}. However, within the EP method 
all contributions are evaluated at zero external momentum, in
contrast to the FD method, which in principle allows non-vanishing
external momentum. Further, the calculation presented in Ref.~\cite{mhiggsEP5} 
is not publicly available as a computer code 
for Higgs-mass calculations. 
Subsequently, another leading three-loop
calculation of \order{\alt\als^2}, depending on the various SUSY mass
hierarchies, has been performed~\cite{mhiggsFD3l},
resulting in the code {\tt H3m} (which
adds the three-loop corrections to the {\tt FeynHiggs} result).
Most recently, a combination of the full one-loop result, supplemented
with leading and subleading two-loop corrections evaluated in the
Feynman-diagrammatic/effective potential method and a resummation of the
leading and subleading logarithmic corrections from the scalar-top
sector has been published~\cite{Mh-logresum} in the latest version of
the code~\fh~\cite{feynhiggs,mhiggslong,mhiggsAEC,mhcMSSMlong,Mh-logresum}.
While previous to this combination the  
remaining theoretical uncertainty on the lightest $\cp$-even Higgs
boson mass had been estimated to be about
$3 \gev$~\cite{mhiggsAEC,PomssmRep}, the combined result was
roughly estimated to yield an uncertainty of 
about  $2 \gev$~\cite{Mh-logresum,ehowp}; 
however, more detailed analyses will be necessary to yield a more solid
result.

In the present paper we calculate the two-loop \order{\alt\als}
corrections to the Higgs boson masses in a mixed
on-shell/\DRbar\ scheme. 
Compared to previously known
results~\cite{mhiggsletter,mhiggslong,mhiggsEP1} we evaluate here
corrections that are proportional to the external momentum of the relevant
Higgs boson self-energies.
These corrections can directly be added to the
corrections included in \fh.
An overview of the relevant sectors and the calculation is given in
\refse{sec:calc}, whereas in \refse{sec:numanal}
we discuss the size and relevance of the new two-loop corrections.
Our conclusions are given in  \refse{sec:conclusions}.


\section{Calculation}
\label{sec:calc}

\subsection{The Higgs-boson sector of the MSSM}
\label{sec:higgs}

The MSSM requires two scalar  doublets, which are conventionally 
written in terms of their components as follows,
\BEA
\cHe &=& \VL \cHe^0 \\[0.5ex] \cHe^- \VR \; = \; \VL v_1 
      + \frac{1}{\sqrt2}(\phi_1^0 - i\chi_1^0) \\[0.5ex] -\phi_1^- \VR\,,
        \non \\
\cHz &=& \VL \cHz^+ \\[0.5ex] \cHz^0 \VR \; = \; \VL \phi_2^+ \\[0.5ex] 
        v_2 + \frac{1}{\sqrt2}(\phi_2^0 + i\chi_2^0) \VR\,.
\label{higgsfeldunrot}
\EEA
The Higgs boson sector can be described with the help of two  
independent parameters (besides the SM gauge couplings), 
conventionally chosen as 
$\tb = v_2/v_1$, the ratio of the two vacuum expectation values, 
and $\MA^2$, the mass of the $\cp$-odd Higgs boson~$A$.
The bilinear part of the Higgs potential leads to 
the tree-level mass matrix for the neutral $\cp$-even Higgs
boson, 
\begin{align}
\label{higgsmassmatrixtree}
M_{\rm Higgs}^{2, {\rm tree}} = \ML \mpe^2 & \mpez^2 \\ 
                           \mpez^2 & \mpz^2 \MR  = 
      \ML \MA^2 \SQb + \MZ^2 \CQb & -(\MA^2 + \MZ^2) \Sbe \Cb \\ 
    -(\MA^2 + \MZ^2) \Sbe \Cb & \MA^2 \CQb + \MZ^2 \SQb \MR ,
\end{align}
in the $(\Pe, \Pz)$ basis and being expressed in terms of the parameters
$M_Z$, $M_A$ and the angle $\beta$. Diagonalization yields the 
tree-level masses $\mhtree$, $\mHtree$.

\bigskip
The higher-order corrected $\cp$-even Higgs boson masses in the
MSSM are obtained  from the corresponding propagators
dressed by their self-energies. 
The inverse propagator matrix in the $(\Pe, \Pz)$ basis is given 
by
\begin{align}
\label{eq:prop}
(\Delta_{\text{Higgs}})^{-1} = -\text{i}
\left( \begin{matrix} 
p^2 - m_{\phi_1}^2 + \hat{\Sigma}_{\phi_1}(p^2) & -m_{\phi_1\phi_2}^2 +\hat{\Sigma}_{\phi_1\phi_2}(p^2)\\ 
-m_{\phi_1\phi_2}^2 +\hat{\Sigma}_{\phi_1\phi_2}(p^2) & p^2 - m_{\phi_2}^2 + \hat{\Sigma}_{\phi_2}(p^2) 
\end{matrix} \right) \text{ ,}
\end{align}
where the $\hat{\Sigma}(p^2)$ denote the renormalized Higgs-boson 
self-energies, $p$ being the external momentum.
The renormalized self-energies can be expressed
through the unrenormalized self-energies, $\Si(p^2)$, and 
counterterms involving renormalization constants 
$\delta m^2$ and $\delta Z$ 
from parameter and field renormalization.
With the self-energies expanded up to two-loop order,
$\hSi  = \hSi^{(1)} + \hSi^{(2)}$,
one has for the $\cp$-even part 
at the $i$-loop level ($i = 1, 2$),
\begin{subequations}
\label{rMSSM:renses_higgssector}
\begin{align}
\seri{\Pe}(p^2)  &=\, \sei{\Pe}(p^2) + \dZi{\Pe}\, (p^2-\mpe^2) - \dmesqi, \\
\seri{\Pz}(p^2)  &=\, \sei{\Pz}(p^2) + \dZi{\Pz}\, (p^2-\mpz^2) - \dmzsqi , \\
\seri{\PePz}(p^2)  &=\, \sei{\PePz}(p^2) - \dZi \PePz \, \mpez^2 - \dmezsqi\,.
\end{align}
\end{subequations}
The counterterms are determined   
by appropriate renormalization conditions 
and are given in the Appendix.

\bigskip
The renormalized self-energies in the $(\Pe, \Pz)$ basis can
be rotated into the physical $(h,H)$ basis where the tree-level 
propagator matrix is diagonal, via
\BE
\ML \ser{HH} & \ser{hH} \\ \ser{hH} & \ser{hh} \MR = 
D(\al) \ML \ser{\Pe} & \ser{\PePz} \\ 
                \ser{\PePz} & \ser{\Pz} \MR D^T(\al)
\label{basis-trans}
\end{equation}
with the matrix
\BE
D(\al) = \ML \Ca & \Sa \\ -\Sa & \Ca \MR\,,
\label{rotmatrix}
\end{equation}
which diagonalizes the tree-level mass matrix~(\ref{higgsmassmatrixtree}).
The $\cp$-even Higgs boson masses are determined by the
poles of the $(h,H)$-propagator matrix. 
This is equivalent to solving the equation
\begin{equation}
\left[p^2 - \mhtree^2 + \hSi_{hh}(p^2) \right]
\left[p^2 - \mHtree^2 + \hSi_{HH}(p^2) \right] -
\left[\hSi_{hH}(p^2)\right]^2 = 0\,~,
\label{eq:proppole}
\end{equation}
yielding the loop-corrected pole masses, $\Mh$ and $\MH$.
Here we use the implementation in the code
\fh~\cite{feynhiggs,mhiggslong,mhiggsAEC,mhcMSSMlong,Mh-logresum},
supplemented by the new momentum dependent \order{\alt\als} corrections,
as described in \ref{sec:fh}.

\begin{figure}[htb!]
\begin{center}
\subfigure[]{\raisebox{0pt}{\includegraphics[width=0.2\textwidth]{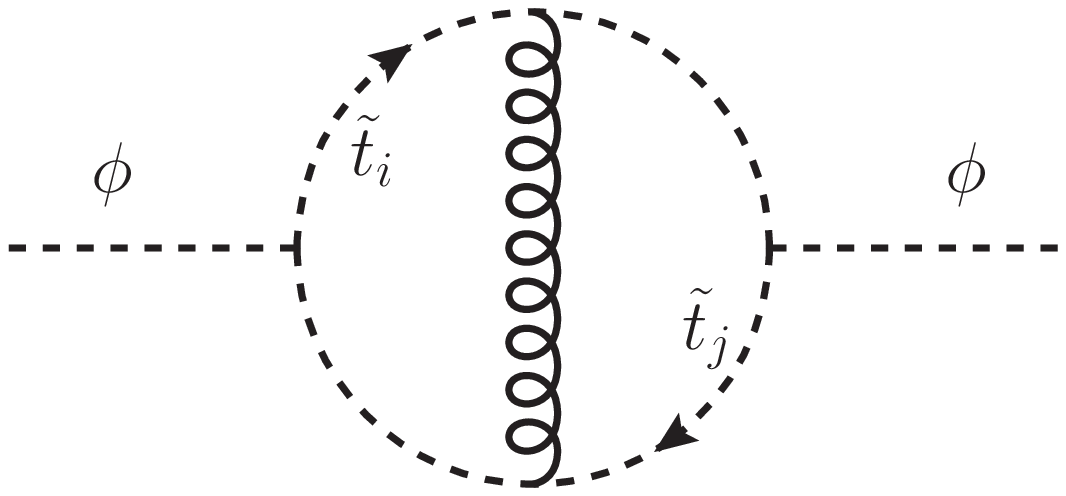}} }
\subfigure[]{\raisebox{0pt}{\includegraphics[width=0.2\textwidth]{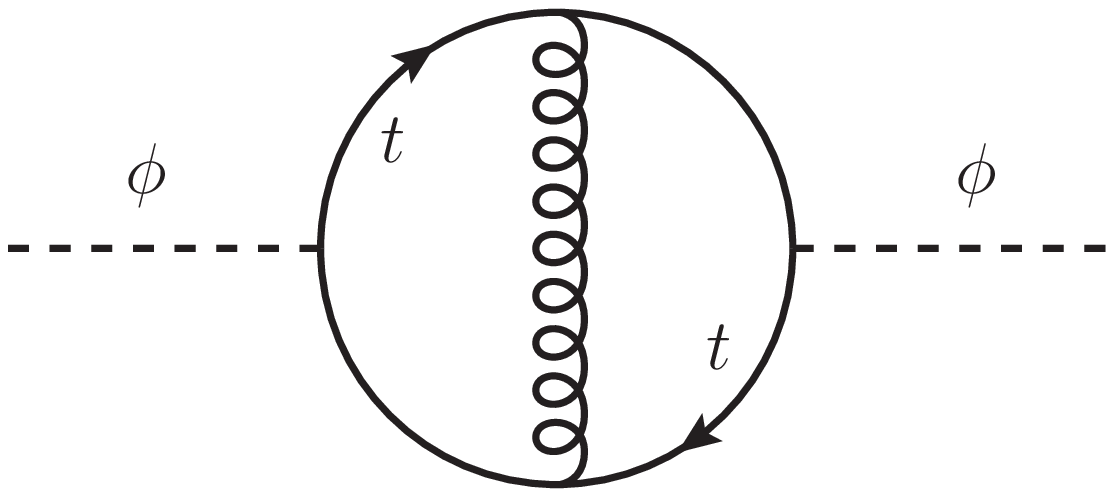}} }
\subfigure[]{\raisebox{0pt}{\includegraphics[width=0.2\textwidth]{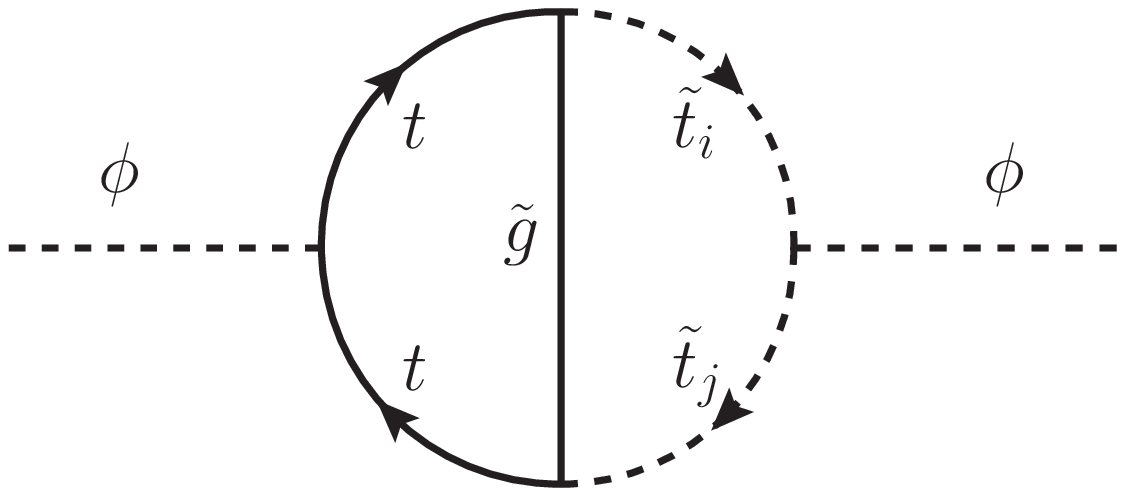}} }
\\%
\subfigure[]{\raisebox{0pt}{\includegraphics[width=0.2\textwidth]{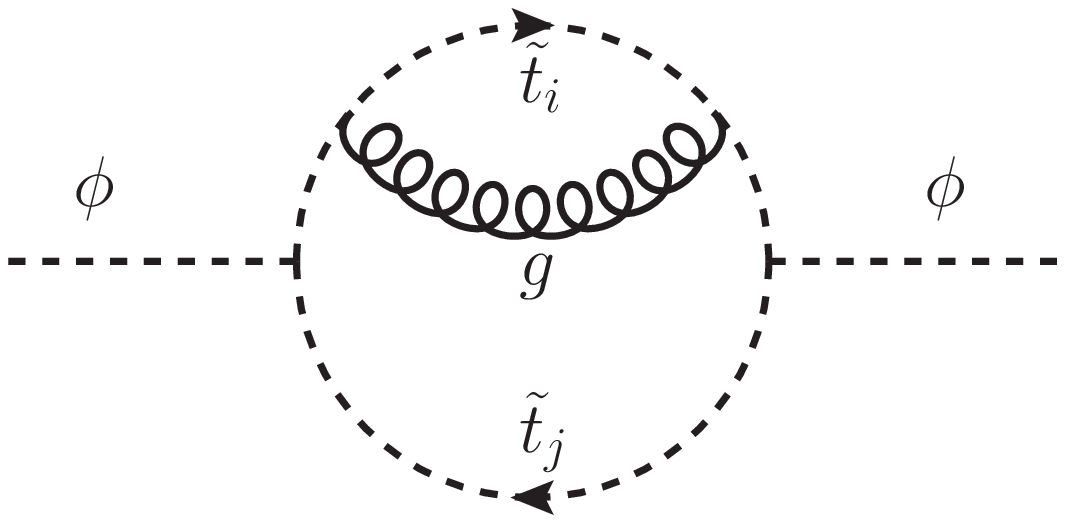}} }
\subfigure[]{\raisebox{0pt}{\includegraphics[width=0.2\textwidth]{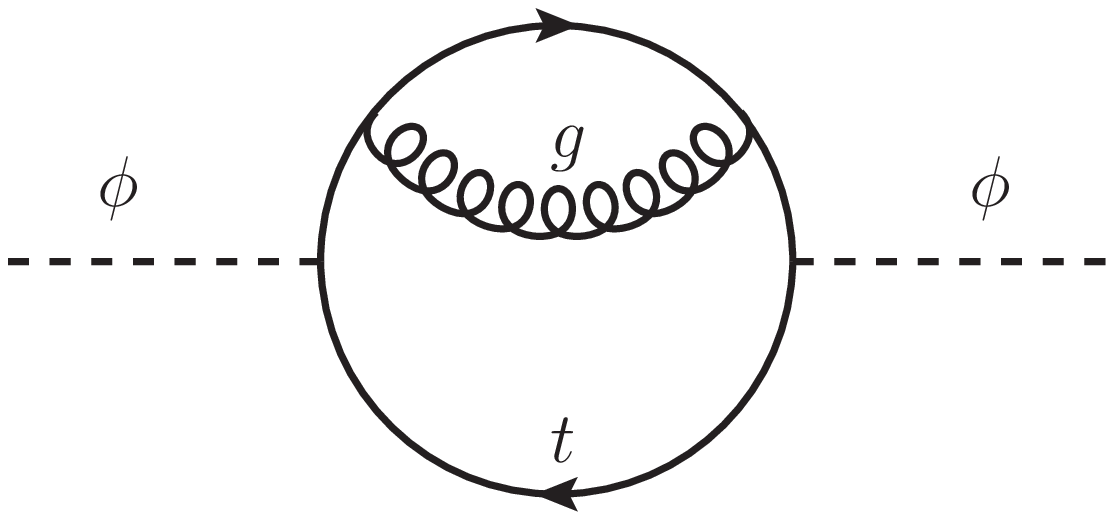}} }
\subfigure[]{\raisebox{0pt}{\includegraphics[width=0.2\textwidth]{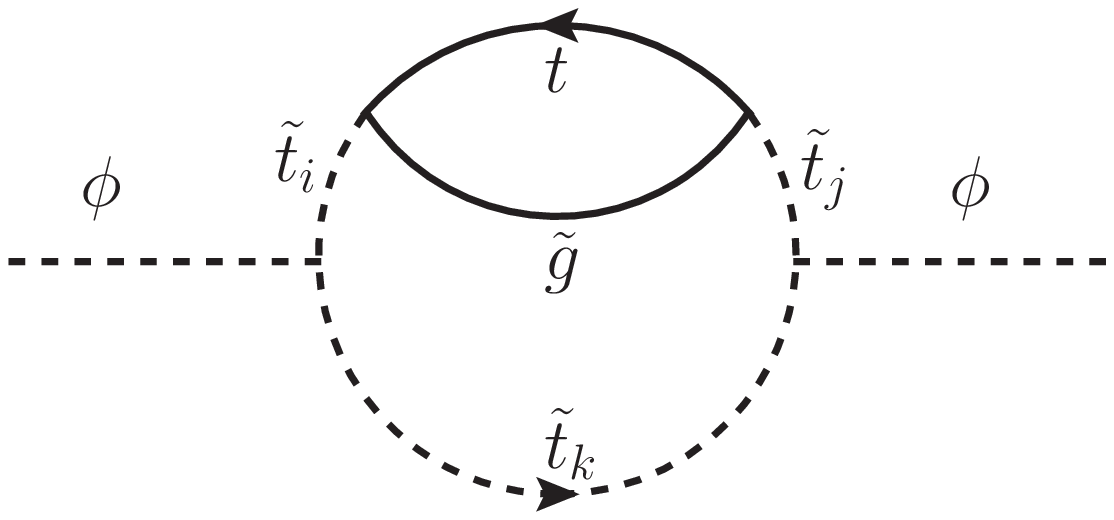}} }\\
\hspace{-20pt}%
\subfigure[]{\raisebox{0pt}{\includegraphics[width=0.2\textwidth]{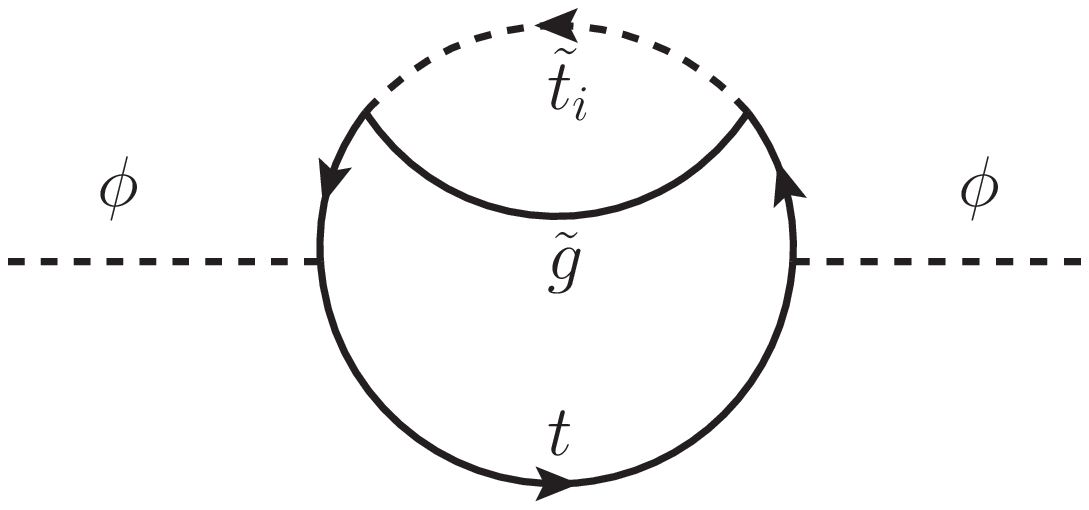}} }\hspace{18pt}
\subfigure[]{\raisebox{0pt}{\includegraphics[width=0.15\textwidth]{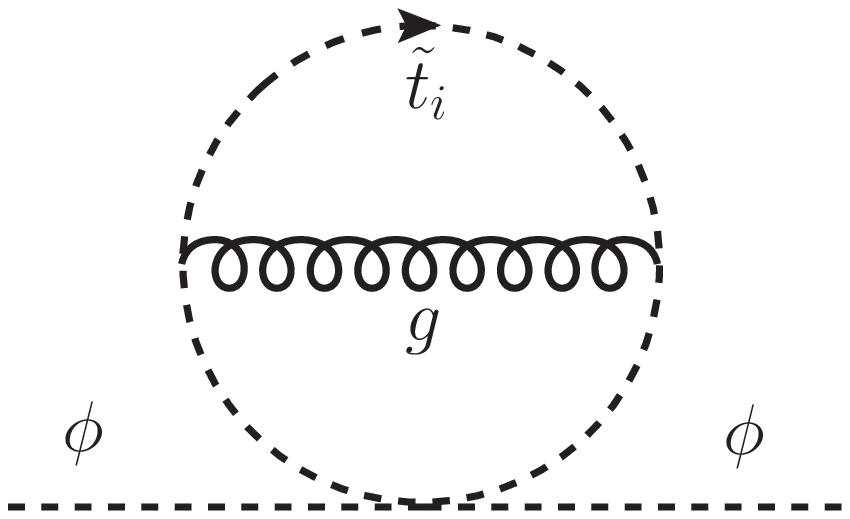}} }\hspace{26pt}
\subfigure[]{\raisebox{0pt}{\includegraphics[width=0.15\textwidth]{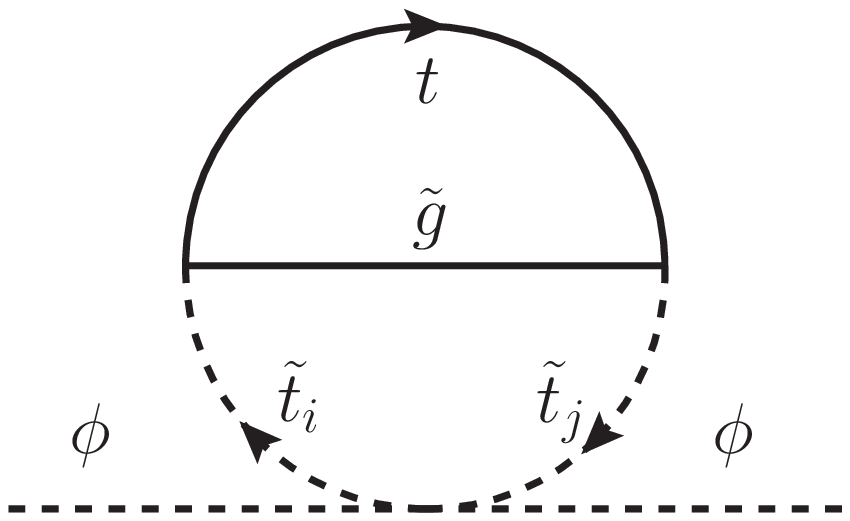}} }\\
\hspace{-4pt}%
\subfigure[]{\raisebox{0pt}{\includegraphics[width=0.15\textwidth]{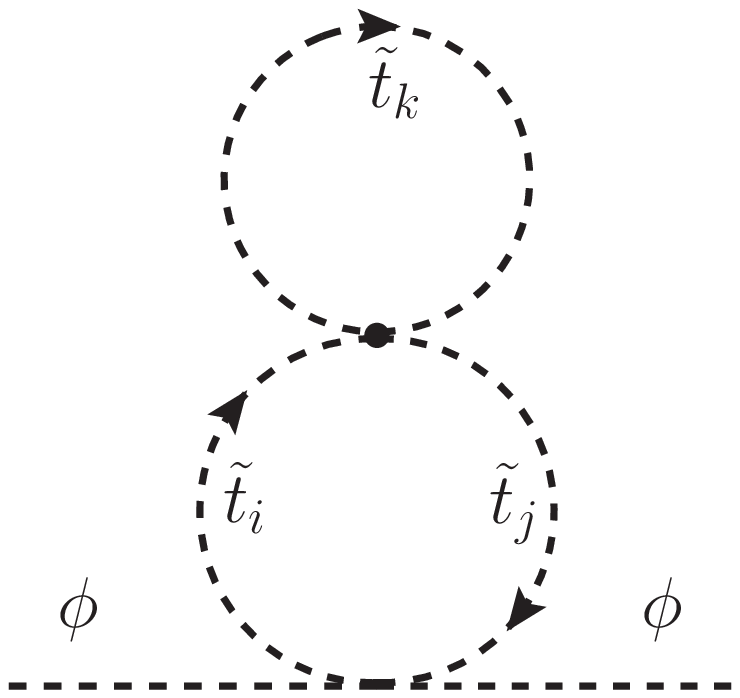}} }\hspace{0pt}
\subfigure[]{\raisebox{-0.5pt}{\includegraphics[width=0.28\textwidth]{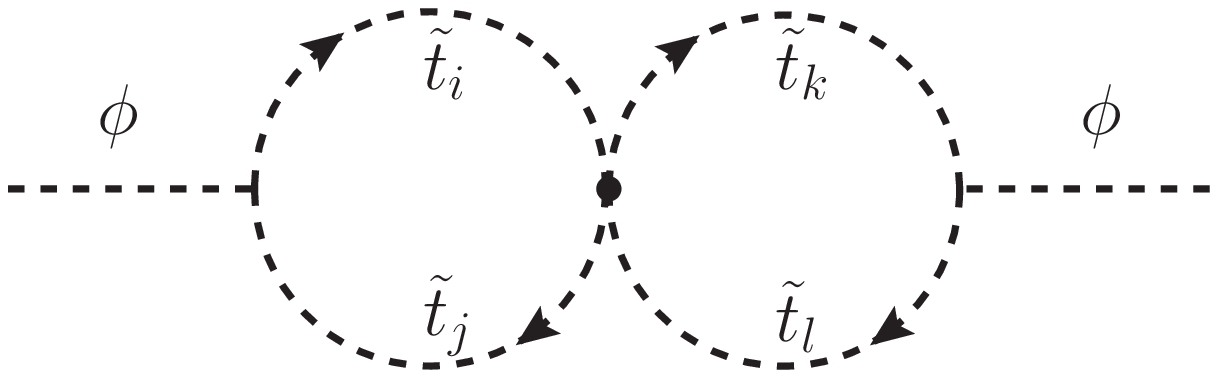}} }\hspace{0pt}
\subfigure[]{\raisebox{2pt}{\includegraphics[width=0.15\textwidth]{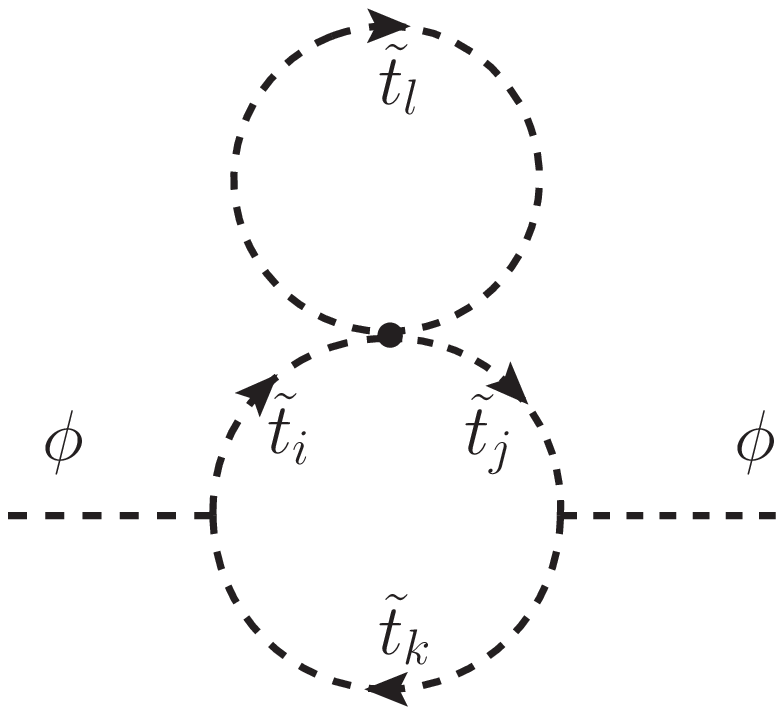}} }
\\%
\subfigure[]{\raisebox{0pt}{\includegraphics[width=0.2\textwidth]{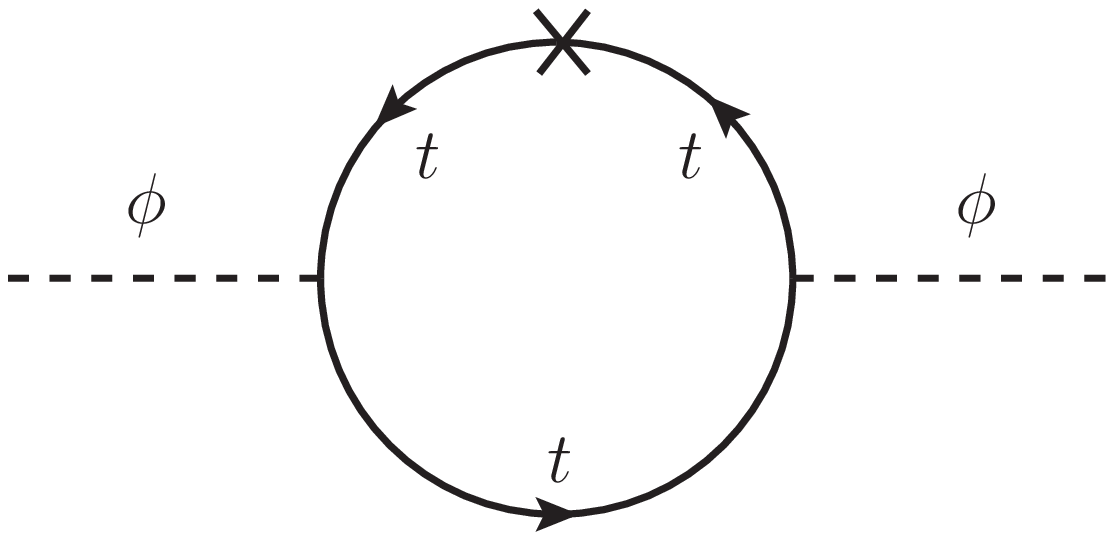}} }
\subfigure[]{\raisebox{0pt}{\includegraphics[width=0.2\textwidth]{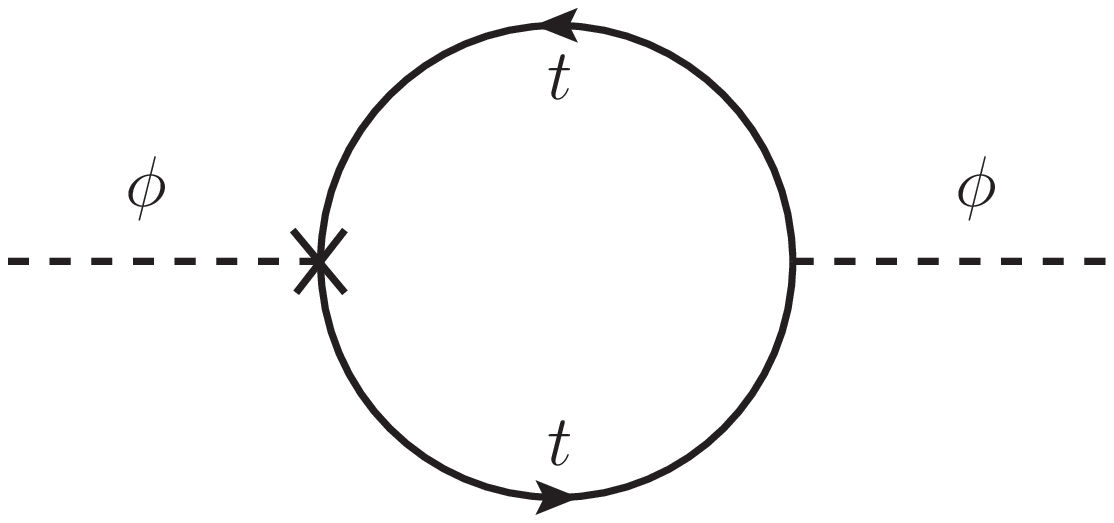}} }
\subfigure[]{\raisebox{-2pt}{\includegraphics[width=0.2\textwidth]{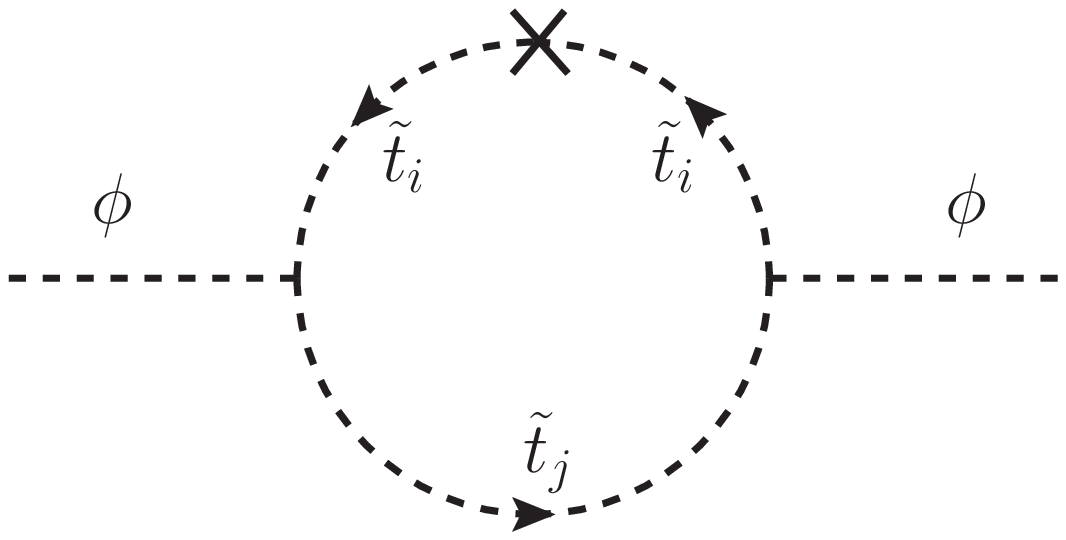}} }\\\hspace{-20pt}
\subfigure[]{\raisebox{0pt}{\includegraphics[width=0.2\textwidth]{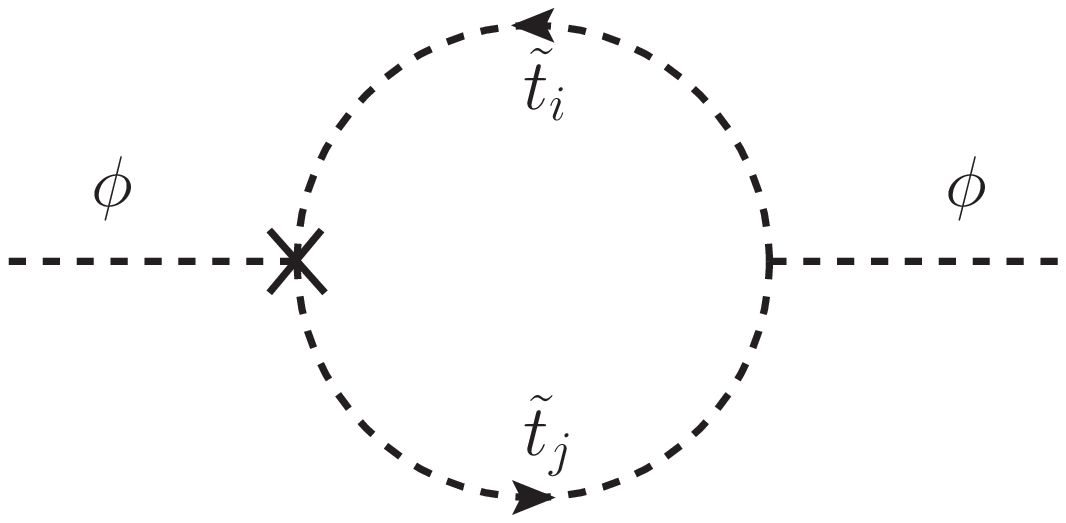}} }\hspace{16pt}
\subfigure[]{\raisebox{0pt}{\includegraphics[width=0.15\textwidth]{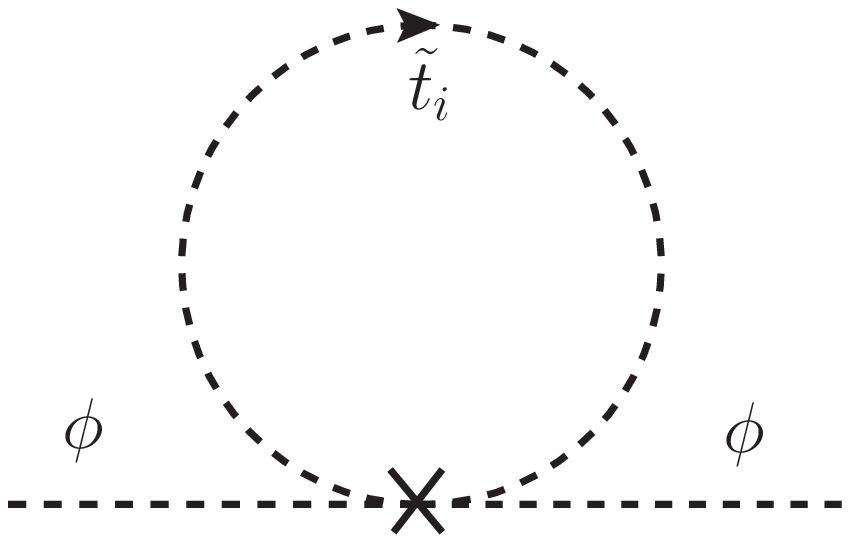}} }\hspace{30pt}
\subfigure[]{\raisebox{0pt}{\includegraphics[width=0.15\textwidth]{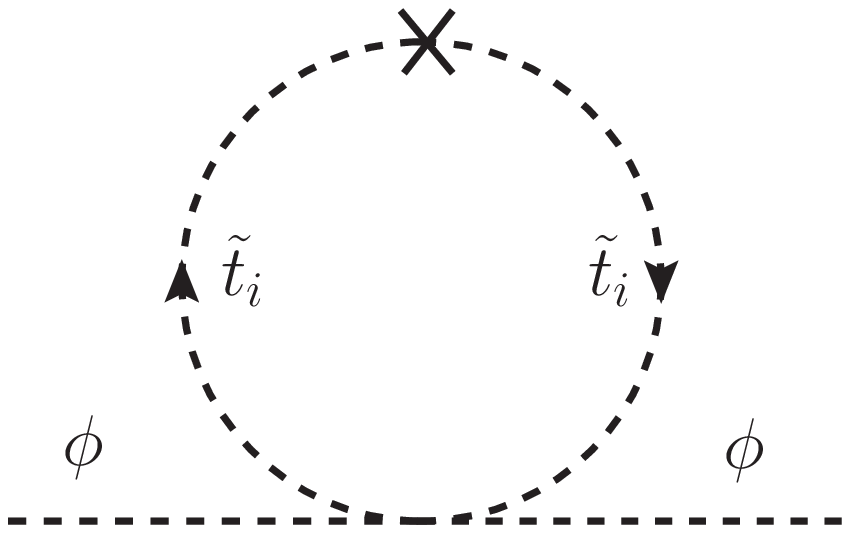}} }
\caption{Generic \twol\ diagrams and diagrams with counterterm
insertions for the Higgs-boson self-energies
($\phi = h, H, A$). }
\label{fig:fd_hHA}
\end{center}
\end{figure}

\begin{figure}[htb!]
\begin{center}
\subfigure[]{\raisebox{0pt}{\includegraphics[width=0.15\textwidth]{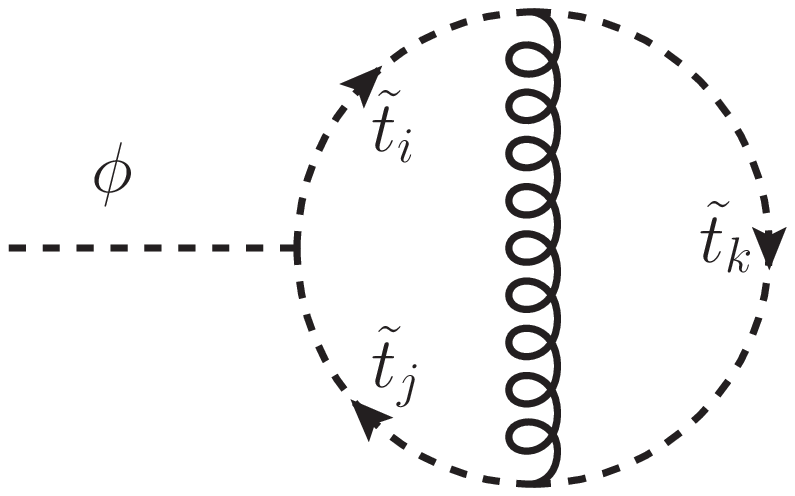}} }\hspace{15pt}
\subfigure[]{\raisebox{0pt}{\includegraphics[width=0.15\textwidth]{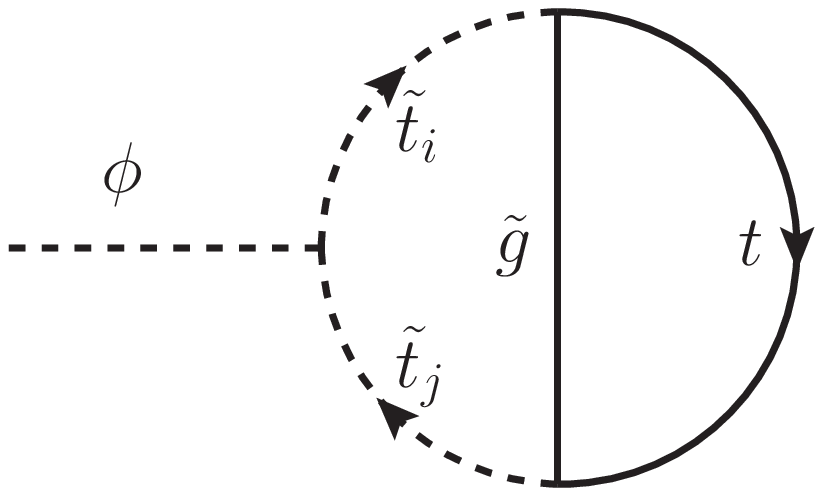}} }\hspace{15pt}
\subfigure[]{\raisebox{7pt}{\includegraphics[width=0.2\textwidth]{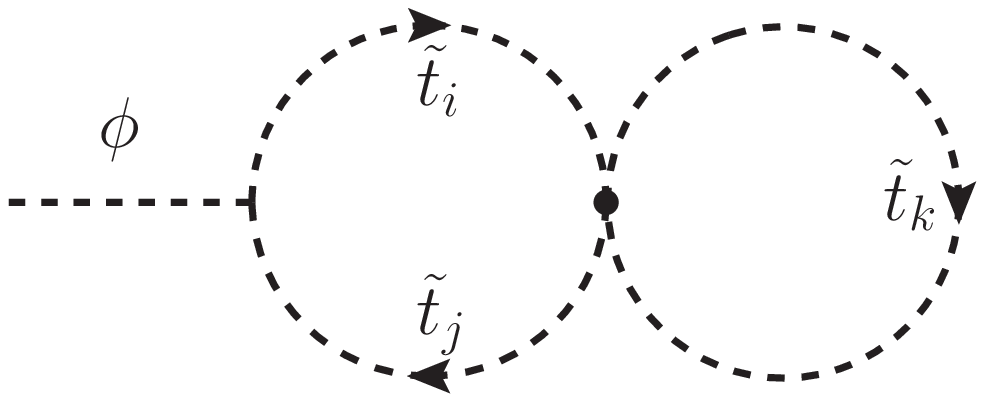}} }\\
\subfigure[]{\raisebox{0pt}{\includegraphics[width=0.15\textwidth]{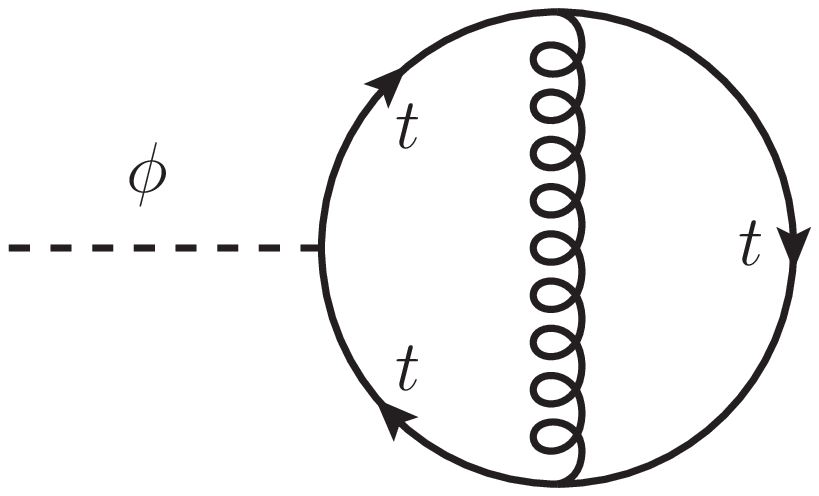}} }\hspace{15pt}
\subfigure[]{\raisebox{0pt}{\includegraphics[width=0.15\textwidth]{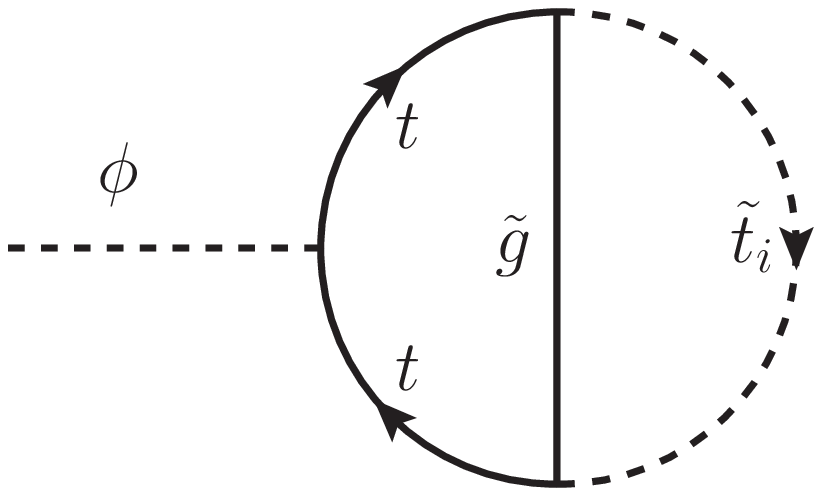}} }\\
\subfigure[]{\raisebox{0pt}{\includegraphics[width=0.15\textwidth]{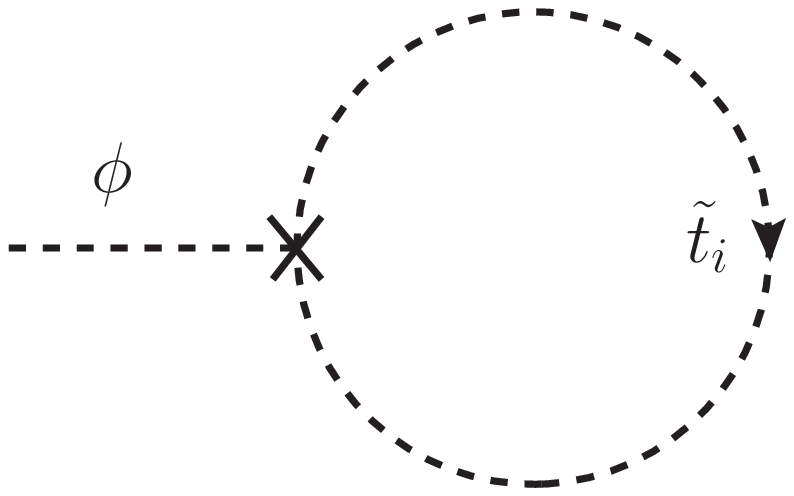}} }\hspace{15pt}
\subfigure[]{\raisebox{0pt}{\includegraphics[width=0.15\textwidth]{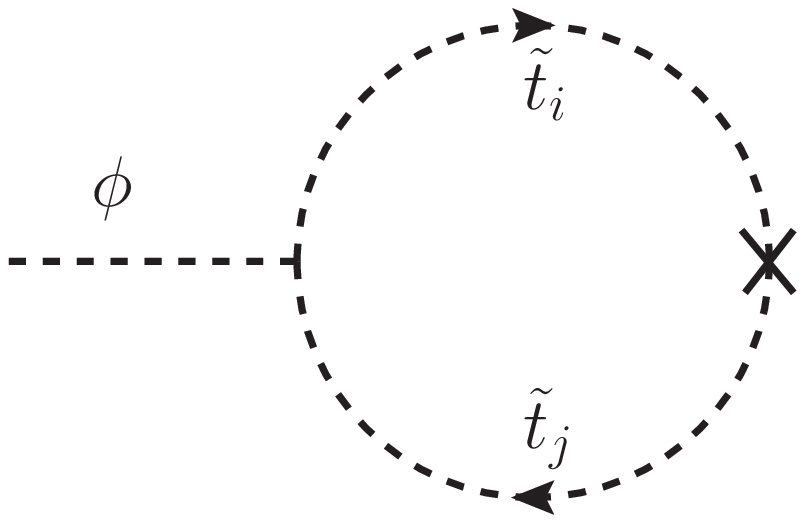}} }\\
\subfigure[]{\raisebox{0pt}{\includegraphics[width=0.15\textwidth]{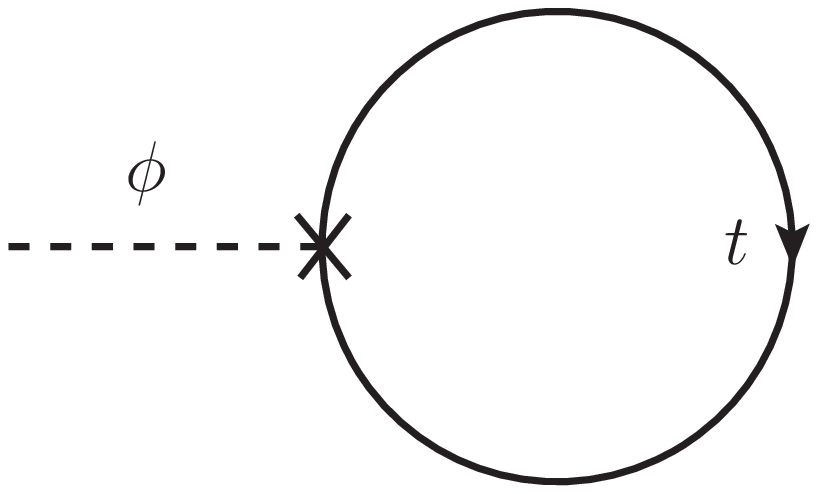}} }\hspace{15pt}
\subfigure[]{\raisebox{0pt}{\includegraphics[width=0.15\textwidth]{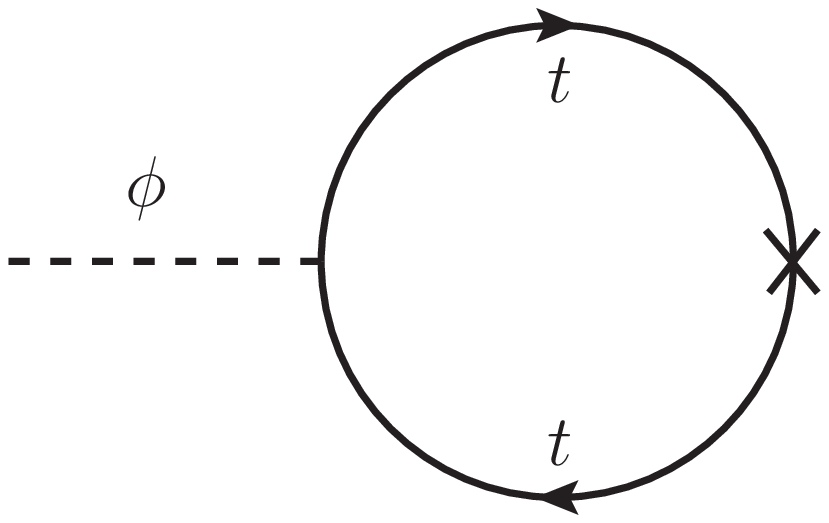}} }
\caption{Generic \twol\ diagrams and diagrams with
  counterterm insertions for the Higgs-boson tadpoles 
($\phi = h,H$; $\;i,j,k = 1,2$).}
\label{fig:fd_TP}
\end{center}
\end{figure}

\begin{figure}[htb!]
\begin{center}
\hspace{-120pt}
\subfigure[]{\raisebox{0pt}{\includegraphics[width=0.2\textwidth]{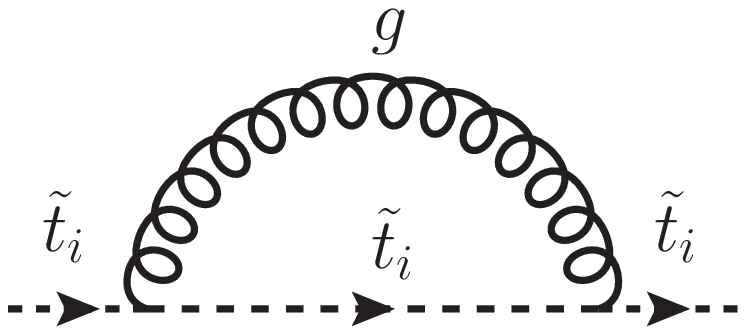}} }\hspace{15pt}
\subfigure[]{\raisebox{0pt}{\includegraphics[width=0.2\textwidth]{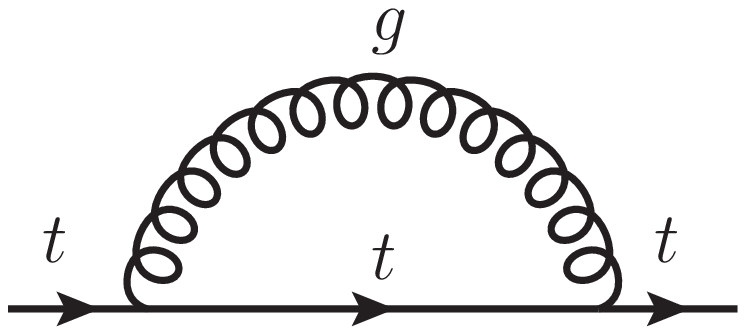}} }\\
\subfigure[]{\raisebox{0pt}{\includegraphics[width=0.2\textwidth]{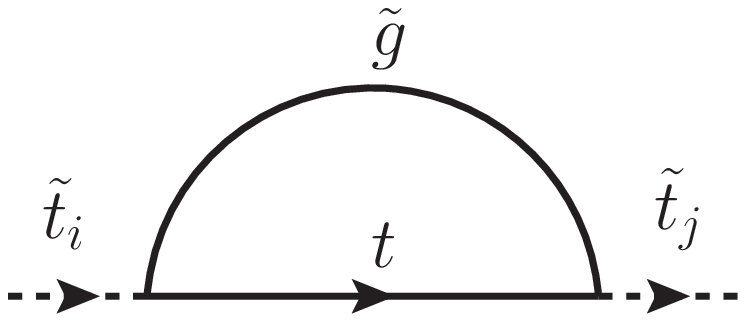}} }\hspace{15pt}
\subfigure[]{\raisebox{0pt}{\includegraphics[width=0.2\textwidth]{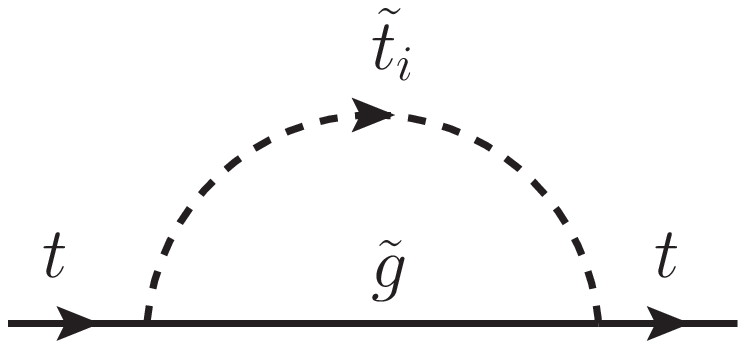}} }\hspace{15pt}
\subfigure[]{\raisebox{1pt}{\includegraphics[width=0.2\textwidth]{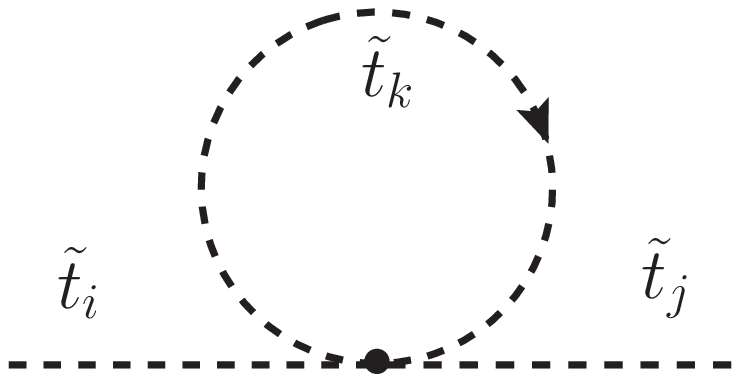}} }
\caption{Generic one-loop diagrams for subrenormalization counterterms, involving top quarks $t$,
top squarks~$\tilde{t}$, gluons $g$ and gluinos $\tilde{g}$  ($i,j,k =1,2$). }
\label{fig:fd_ctis}
\end{center}
\end{figure}

\smallskip
Our calculation is performed in the Feynman-diagrammatic (FD) approach. To
arrive at expressions for the unrenormalized self-energies and tadpoles
at \order{\alt\als}, the evaluation of genuine two-loop diagrams
and one-loop graphs with counterterm insertions is required.
Example diagrams for the neutral Higgs-boson self-energies are shown
in \reffi{fig:fd_hHA}, and for the tadpoles in~\reffi{fig:fd_TP}.
For the counterterm insertions, described in subsection~\ref{sec:stop},
one-loop diagrams with external top quarks/squarks have 
to be evaluated as well, as displayed in~\reffi{fig:fd_ctis}. 
The complete set of contributing Feynman diagrams 
has been generated with the
program {\tt FeynArts}~\cite{feynarts} (using the model file including
counterterms from \citere{mssmct}),
tensor reduction and the evaluation of traces was done with 
support from the programs~{\tt FormCalc}~\cite{formcalc} and 
{\tt TwoCalc}~\cite{twocalc},
yielding algebraic expressions in terms of 
the scalar one-loop functions $A_0, B_0$~\cite{oneloop}, the 
massive vacuum two-loop functions~\cite{Davydychev:1992mt}, 
and two-loop integrals which depend on the external momentum.
These integrals have been evaluated with the program 
\sd~\cite{Carter:2010hi,Borowka:2012yc}, see subsection~\ref{sec:sd}.

\clearpage


\subsection{The scalar-top sector of the MSSM}
\label{sec:stop}

The bilinear part of the top-squark Lagrangian,
\begin{align}
\cL_{\Stop, \text{mass}} &= - \begin{pmatrix}
{{\tilde{t}}_{L}}^{\dagger}, {{\tilde{t}}_{R}}^{\dagger} \end{pmatrix}
\matr{M}_{\tilde{t}}\begin{pmatrix}{\tilde{t}}_{L}\\{\tilde{t}}_{R}
\end{pmatrix} \,,
\end{align}
contains the stop mass matrix $\matr{M}_{\tilde{t}}$,
given by 
\begin{align}\label{Sfermionmassenmatrix}
\matr{M}_{\tilde{t}} 
&= \begin{pmatrix}  
 \MstL^2 + \mt^2 + \MZ^2 \CZb \, (T_t^3 - Q_t \sw^2) & 
 \mt \Xt \\[.2em]
 \mt \Xt &
 \MstR^2 + \mt^2 + \MZ^2 \CZb \, Q_t \, \sw^2
\end{pmatrix}\,, \\
{\rm with} &\mbox{} \non \\
\Xt &= \At - \mu\,\CTb\,.
\end{align}
$Q_t$ and $T_t^3$ denote the charge and isospin of the top quark, 
$\At$ is the trilinear coupling between the Higgs bosons and the scalar
tops, and $\mu$ is the Higgsino mass parameter.
Below we use $\msusy := \MstL = \MstR$ for our numerical evaluation.
However, the analytical calculation has been performed for arbitrary 
$\MstL$ and $\MstR$. 
$\matr{M}_{\tilde{t}}$  can be diagonalized with the help of a unitary
transformation matrix ${\matr{U}}_{\tilde{t}}$, parametrized
by a mixing angle $ {\theta}_{\tilde{t}}$, 
to provide the eigenvalues $\mste^2$ and $\mstz^2$
as the squares of the two on-shell top-squark masses.  

\medskip
For the evaluation of the \order{\alt\als} two-loop contributions
to the self-energies and tadpoles of the Higgs sector,
renormalization of the top/stop sector at \order{\als} is required,
giving rise to the counterterms for one-loop subrenormalization
(see \reffis{fig:fd_hHA},\ref{fig:fd_TP}). 
We follow the renormalization at the one-loop level given in 
\citeres{mhiggsFD2,hr,SbotRen,Stop2decay}, where details can be
found. In the context of this paper, we only want to emphasize
that on-shell (OS) renormalization is performed
for the top-quark mass as well as for the  scalar-top masses.
This is different from the approach pursued,
for example, in~\citere{mhiggs2lp2}, where a \DRbar\ 
renormalization has been employed. Using the OS scheme allows us to
consistently combine our new correction terms with the 
hitherto available self-energies included in \fh.

\medskip
Finally, at \order{\alt\als}, gluinos appear as
virtual particles only at the two-loop level (hence, no renormalization
for the gluinos is needed). The corresponding 
soft-breaking gluino mass parameter $M_3$ determines the gluino mass, 
$\Mgl = M_3$.


\subsection{The program \sd}
\label{sec:sd}

\begin{figure}[htb!]
\begin{center}
\begin{minipage}{0.99\textwidth}
\includegraphics[width=0.23\textwidth]{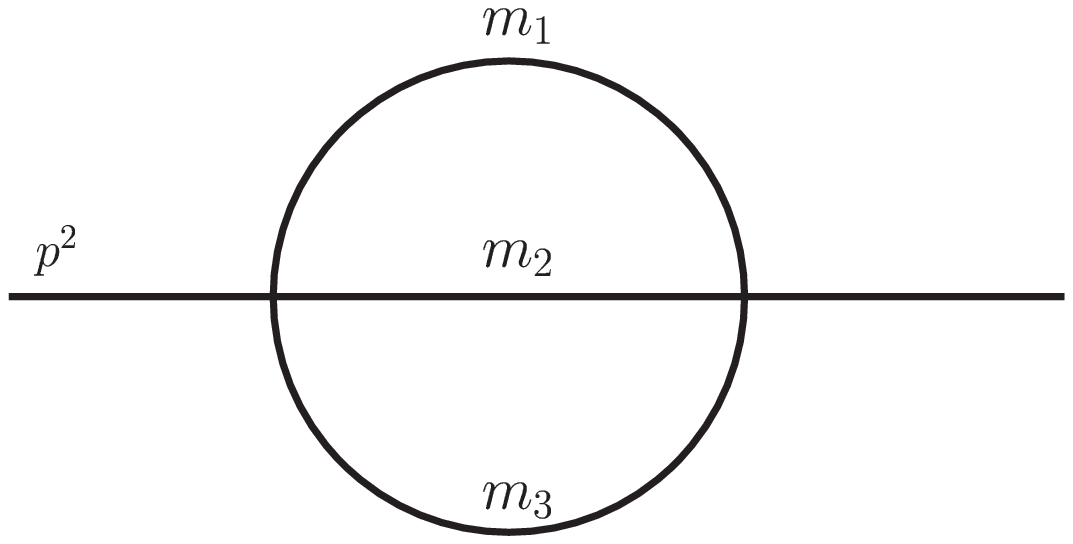}\hspace{1pt}
\includegraphics[width=0.23\textwidth]{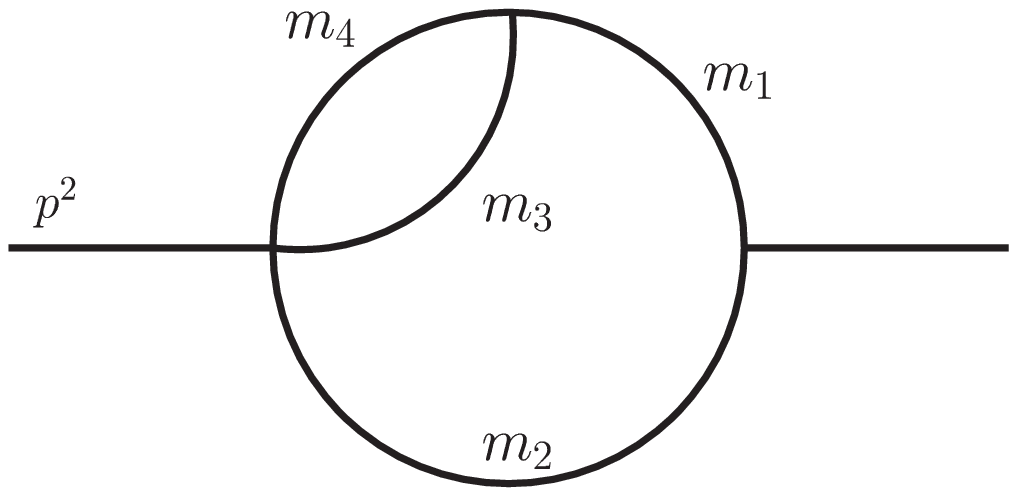}\hspace{1pt}
\includegraphics[width=0.23\textwidth]{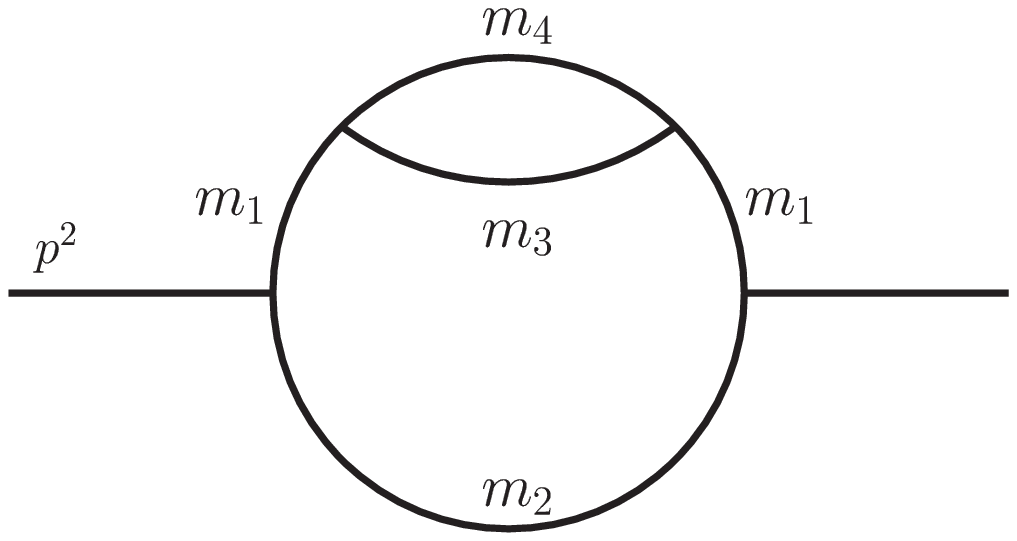}\hspace{1pt}
\includegraphics[width=0.23\textwidth]{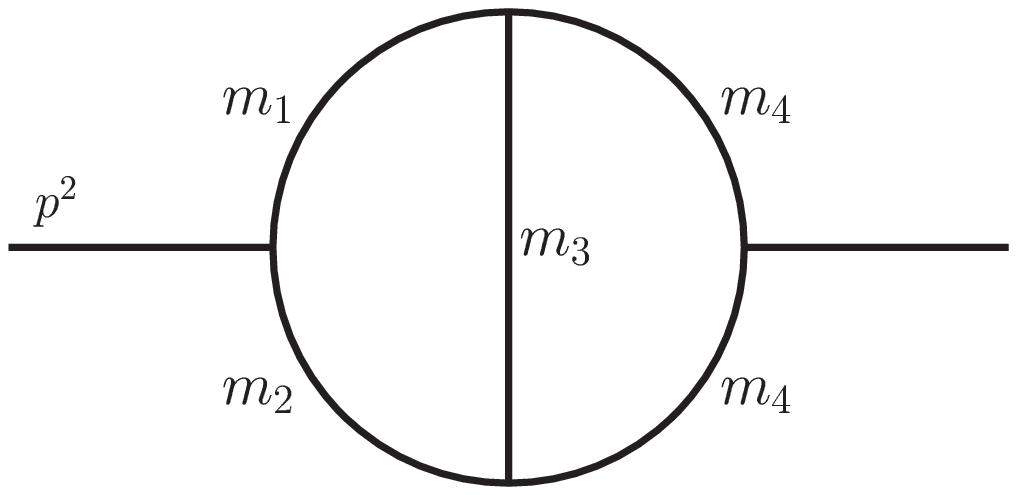}\vspace{-15pt}\\
\begin{center} 
\hspace*{-15pt}$T_{234}$ \hspace{85pt} $T_{1234}$\hspace{85pt} $T_{11234}$\hspace{80pt} $T_{12345}$\vspace{10pt}\\
\end{center} 
\end{minipage}
\caption{Topologies which have been calculated numerically using \sd.}
\label{fig:Tintegrals}
\end{center}
\end{figure}

The calculation of the momentum-dependent two-loop corrections to the
Higgs-boson masses at order \order{\alt\als} involves two-loop two-point
functions with up to four different masses, in addition to the mass
scale given by the external momentum $p^2$. For two-loop diagrams of
propagator type, analytical results in four space-time dimensions are
known only sparsely if different masses are occurring in the
loops~\cite{Broadhurst:1987ei,Davydychev:1992mt,Davydychev:1993pg,Scharf:1993ds,Berends:1994ed,Bauberger:1994hx, Bauberger:1994zz,Laporta:2004rb,Remiddi:2013joa}. 
The integrals which are lacking analytical results 
can be classified into four 
different topologies, shown in Figure~\ref{fig:Tintegrals}. 
We have calculated these integrals  numerically using 
the program \sd~\cite{Carter:2010hi,Borowka:2012yc}, where 
up to four different masses in 34 different mass configurations needed 
to be considered, with differences in the kinematic invariants of several orders of magnitude.

The program \sd{} is a publicly available tool~\cite{hepforge} 
to calculate multi-loop integrals numerically. 
Dimensionally regulated poles are factorized by sector decomposition as 
described in~\citeres{Binoth:2000ps,Heinrich:2008si}, while kinematic thresholds are handled by 
a deformation of the integration contour into the complex plane, as described \eg 
in~\citere{Borowka:2012yc}. The numerical integration is done using the {\sc Cuba} 
library~\cite{Hahn:2004fe}.

The program  has also been extended 
to be able to calculate tensor integrals of any rank~\cite{Borowka:2013cma}, 
and to process efficiently the evaluation of large ranges of kinematic 
points using the ``{\tt multinumerics}" feature of the program, 
which is of particular importance for the calculation presented here.
This feature allows to produce input files for large sets of kinematic points 
automatically, and to process the evaluation of these points in parallel 
if several cores or a cluster are available, without repeating the algebraic part
of the sector decomposition, which can be done once and for all.
The evaluation of a single phase space point for the most complicated topology, 
to reach a relative accuracy of at least $10^{-5}$,  
ranges between 0.01 and 100 seconds on an Intel core i7 processor, 
where the larger timings are for points very close to a kinematic threshold.


\subsection{Evaluation and implementation in the program \fh}
\label{sec:fh}

The resulting new contributions to the neutral $\cp$-even Higgs-boson 
self-energies, containing all momentum-dependent and additional constant 
terms, are assigned to the differences
\begin{equation}
\label{eq:DeltaSE}
\De\ser{ab}^{(2)}(p^2) = \ser{ab}^{(2)}(p^2) - \tilde\Sigma_{ab}^{(2)}(0)\,,
\qquad
ab = \{HH,hH,hh\}\,.
\end{equation}
Note the tilde (not hat) on $\tilde\Sigma^{(2)}(0)$ which signifies that 
not only the self-energies are evaluated at zero external momentum but
also the corresponding counterterms,
following \citeres{mhiggsletter,mhiggslong}.
A finite shift $\De\hat{\Sigma}^{(2)}  (0)$
therefore remains in the limit $p^2\to 0$ 
due to $\dMAsqt = \re\se{AA}^{(2)}(\MA^2)$ being computed at $p^2=\MA^2$ 
in $\hat\Sigma^{(2)}$, but at $p^2=0$ in $\tilde\Sigma^{(2)}$; for details 
see \refeqs{rMSSM:mass_osdefinition} and (\ref{masscounterterms}) 
in the Appendix.

\bigskip
The numerical evaluation to derive the physical masses for $h,H$  as the
poles (real parts) of the dressed propagators proceeds on the basis of~\refeq{eq:proppole}
in an iterative way.
\begin{itemize}
\item[$\bullet$]
In a first step, the squared masses $M_{h,0}^2, M_{H,0}^2$ are determined 
by solving \refeq{eq:proppole} excluding the new terms $\De\hat\Sigma_{ab}^{(2)}(p^2)$
from the self-energies.
\item[$\bullet$]
In a second step, the shifts $\De\hat\Sigma^{(2)}_{ab}(M_{h,0}^2) \equiv c_{ab}^h$ and
$\De\hat\Sigma^{(2)}_{ab}(M_{H,0}^2) \equiv c_{ab}^H$ are calculated and 
added as constants to the self-energies in \refeq{eq:proppole},
$\ser{ab}(p^2) \to \ser{ab}(p^2) + c_{ab}^{h(H)}$.
\item[$\bullet$]
In the third step, \refeq{eq:proppole} is solved again, now including the
constant shifts $c_{ab}^{h(H)}$ in the self-energies, to deliver the 
refined masses $\Mh$ (with $c_{ab}^h)$ and $\MH$ (with $c_{ab}^H$). 
\end{itemize}
This procedure can be repeated for improving the accuracy; numerically
it turns out that going beyond the first iteration yields only marginal changes.

\bigskip 
The corrections of \refeq{eq:DeltaSE}  are  incorporated in \fh{} by 
the following recipe, which is more general and in principle
applicable also to the case of the complex MSSM with $\cp$ violation.
\begin{enumerate}
\item Determine Higgs masses $M_{h_{i},0}$ 
without the momentum-dependent terms of \refeq{eq:DeltaSE}; 
the index $i=1,\dots,4$ enumerates 
the masses of $h, H, A, H^\pm$ in the real MSSM. 
This is done by invoking the \fh\ mass-finder.

\item Compute the shifts 
 $c_{ab}^{h_k} = \De\ser{ab}^{(2)} (M_{h_k,0}^2)$  with $a,b,h_k = {h,H}$.

\item Run \fh' mass-finder again including the $c_{ab}^{h_k}$
as constant shifts in the self-energies
to determine the refined Higgs masses $M_{h}$ and $M_{H}$.
\end{enumerate}
This procedure could conceivably be iterated until full self-consistency 
is reached; yet the resulting mass improvements turn out to be too small 
to justify extra CPU time.

\bigskip
On the technical side we added an interface for an external program to 
\fh{} which exports relevant model parameters to the external program's 
environment, currently: 
\begin{tabbing}
\texttt{FHscalefactor} \qquad\= ren. scale multiplicator, 
\qquad\qquad \= 	
	\texttt{FHTB}	\qquad	\= $\tan\beta$, \\
         \texttt{FHAlfasMT}  \> $\alpha_s(\mt)$, \>
      \texttt{FHGF}	\qquad	\= $G_F$, \\
      \texttt{FHMHiggs2$i$} \qquad\> $M_{h_i,0}^2$, $i = 1\dots 4$
      , \>
     \texttt{FHMSt$i$}		\> $m_{\tilde t,i}$, $i = 1, 2$, \\
     \texttt{FH\{Re,Im\}USt1$i$} \> $U_{\tilde t,1i}$, $i = 1, 2$, \>
       \texttt{FHMGl}		\> $m_{\tilde g}$ , \\
     \texttt{FH\{Re,Im\}MUE}	\> $\mu$ , \>
     \texttt{FHMA0}		\> $M_A$, 
\end{tabbing}
where the $U_{\tilde t,1i}$ denote the elements of the stop mixing 
matrix, $\alpha_s(\mt)$ the running strong coupling 
at the scale $\mt$, and $G_F$  
the Fermi constant. The renormalization 
scale is defined within \fh\
as $\mu_R = m_t \cdot \texttt{FHscalefactor}$. 
Invocation of the external program is switched on by providing its path 
in the environment variable \texttt{FHEXTSE}.  The program is executed 
from inside a temporary directory which is afterwards removed.

\smallskip
The output (stdout) is scanned for lines of the form 
`\textit{se}@$m$ $c_r$ $c_i$'
which specify the correction $c_r + \mathrm{i} c_i$ [with
$c_r = {\rm Re}(c_{ab}^{h_k}), \; c_i = {\rm Im}(c_{ab}^{h_k})$]
to self-energy 
\textit{se} in the computation of mass $m$, where $m$ is one of 
\texttt{Mh0}, \texttt{MHH}, \texttt{MA0}, \texttt{MHp}, and \textit{se} 
is one of \texttt{h0h0}, \texttt{HHHH}, \texttt{A0A0}, \texttt{HmHp}, 
\texttt{h0HH}, \texttt{h0A0}, \texttt{HHA0}, \texttt{G0G0}, 
\texttt{h0G0}, \texttt{HHG0}, \texttt{A0G0}, \texttt{GmGp}, 
\texttt{HmGp}, \texttt{F1F1}, \texttt{F2F2}, \texttt{F1F2}.  The latter 
three, if given, substitute
\begin{subequations}
\begin{align}
\mathtt{HHHH} &= \cos^2\alpha\,\mathtt{F1F1} +
                 \sin^2\alpha\,\mathtt{F2F2} +
                 \sin 2\alpha\,\mathtt{F1F2}\,, \\
\mathtt{h0h0} &= \sin^2\alpha\,\mathtt{F1F1} + 
                 \cos^2\alpha\,\mathtt{F2F2} -
                 \sin 2\alpha\,\mathtt{F1F2}\,, \\
\mathtt{h0HH} &= \cos 2\alpha\,\mathtt{F1F2} +
                 \tfrac 12\sin 2\alpha\,(\mathtt{F2F2} - \mathtt{F1F1})\,,
\end{align}
\end{subequations}
where $\alpha$ is the tree-level $2\times 2$ neutral-Higgs mixing angle
in~\refeq{rotmatrix}.
Self-energies not given are assumed zero.

\smallskip
The zero-momentum contributions $\tilde\Sigma_{ab}^{(2)}(0)$, $ab = 
\{HH,hH,hh\}$, are subtracted if the output of the external program contains one or more of 
`\texttt{sub asat}', `\texttt{sub atat}', `\texttt{sub asab}', 
`\texttt{sub atab}' for the $\alpha_s\alpha_t$, $\alpha_t^2$, 
$\alpha_s\alpha_b$, and $\alpha_t\alpha_b$ contributions, respectively.
All other lines in the output are ignored.



\section{Numerical results}
\label{sec:numanal}

We show results for the subtracted two-loop self-energies
$\De\ser{ab}^{(2)}(p^2)$ given in
\refeq{eq:DeltaSE}, as well as for the mass shifts
\begin{align}
\De \Mh = \Mh - M_{h,0}, \quad \De \MH = \MH - M_{H,0}
\end{align} 
i.e.\ the difference in the physical Higgs-boson masses
evaluated including and excluding the newly obtained momentum-dependent
two-loop corrections.
This quantity, in particular $\De\Mh$ for the light
$\cp$-even Higgs boson, can directly be compared with the current
experimental uncertainty as well as with the anticipated future ILC
accuracy of~\cite{dbd}, 
\begin{align}
\de\Mh^{\rm exp, ILC} \lsim 0.05 \gev~.
\label{dMhILC}
\end{align}
The results are obtained for two different scenarios, 
varying  parameters like $\tb,\MA,\Mgl$, and illustrate the impact 
of these parameters via the new two-loop corrections on
the neutral $\cp$-even Higgs boson masses, $\Mh$ and $\MH$.
The corresponding renormalization scale, $\mudim$, is set to 
$\mudim = \mt$ in all numerical evaluations. 
The scale uncertainties are expected to be much smaller than the 
parametric uncertainties due to variations of parameters like 
$\tb,\MA,\Mgl,m_{\tilde{t}}$.


\subsection{Scenario 1: \boldmath{\mhmax\ }}

\begin{figure}[ht!]
\centering
\includegraphics[width=0.49\textwidth]{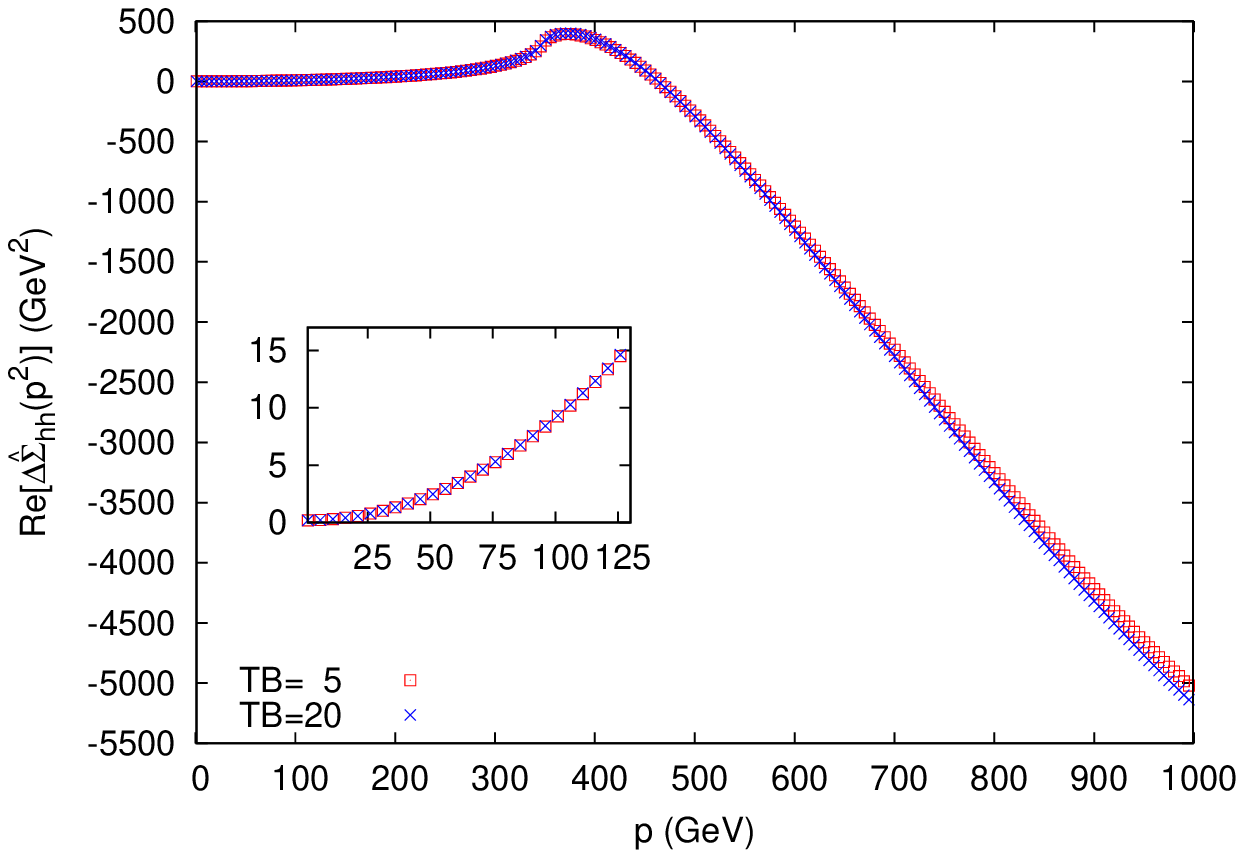}
\includegraphics[width=0.49\textwidth]{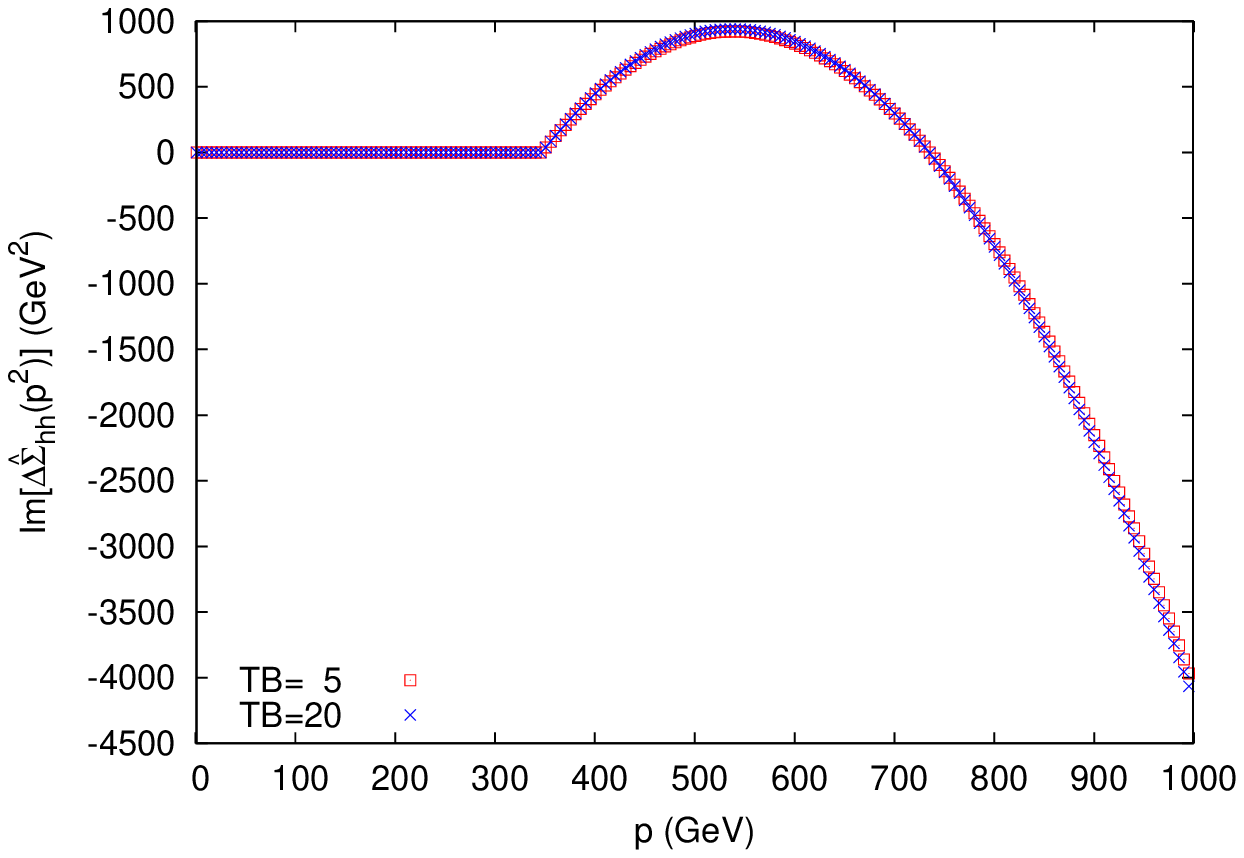}
\\[2em]
\includegraphics[width=0.49\textwidth]{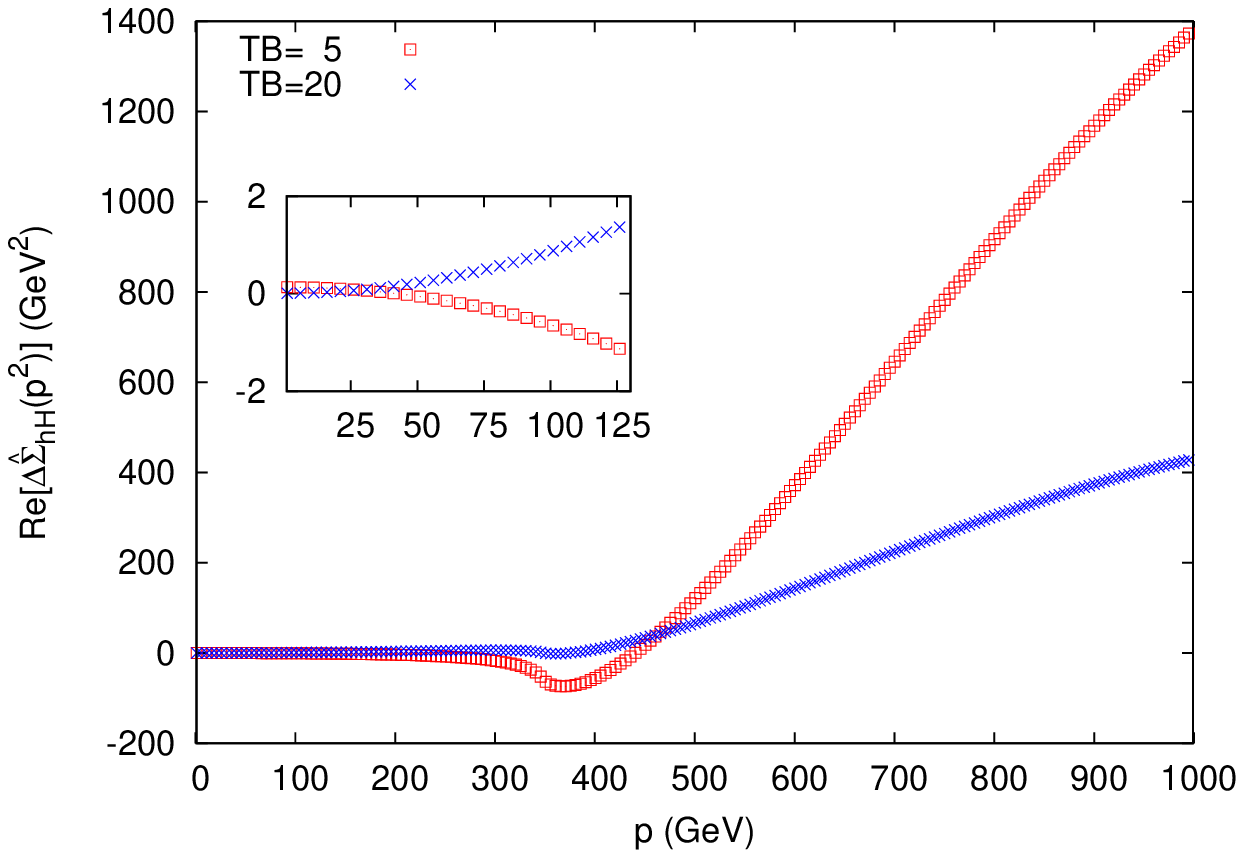}
\includegraphics[width=0.49\textwidth]{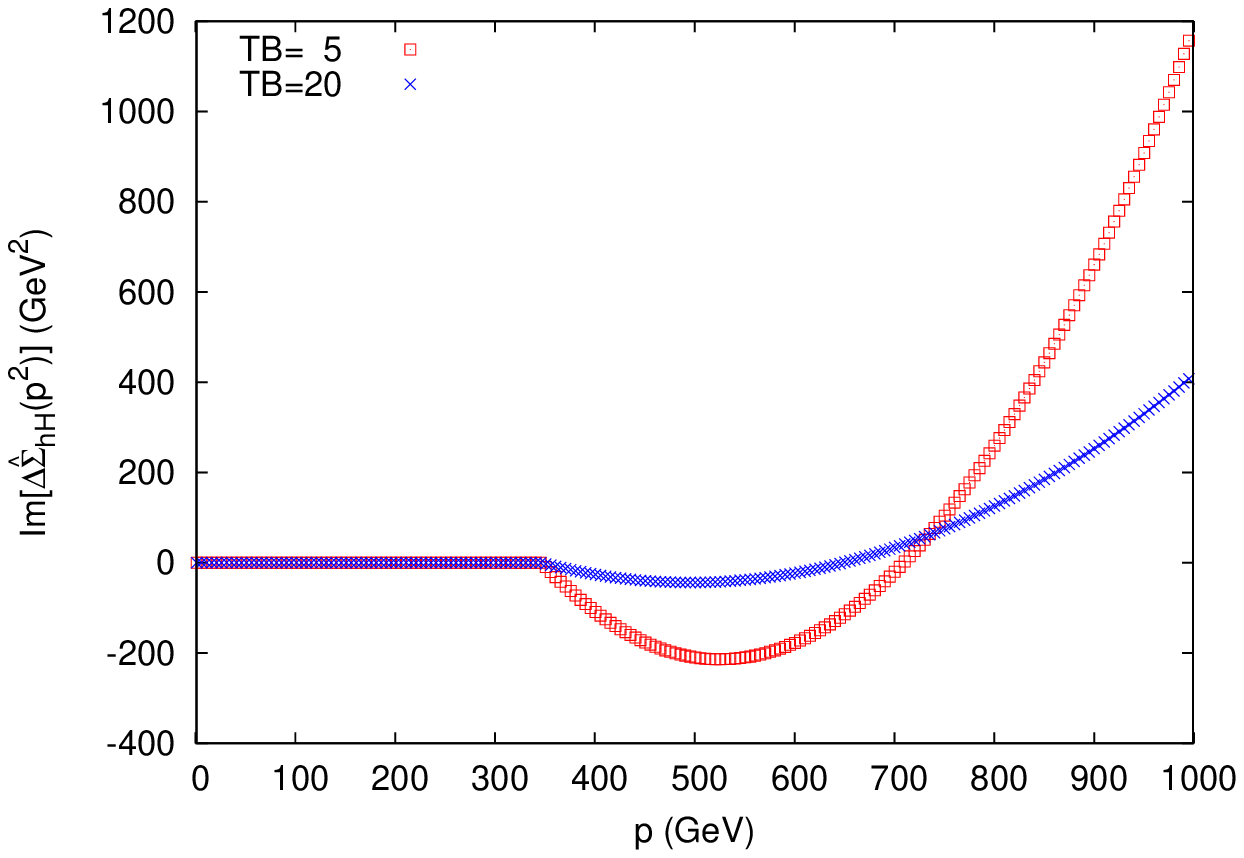}
\\[2em]
\includegraphics[width=0.49\textwidth]{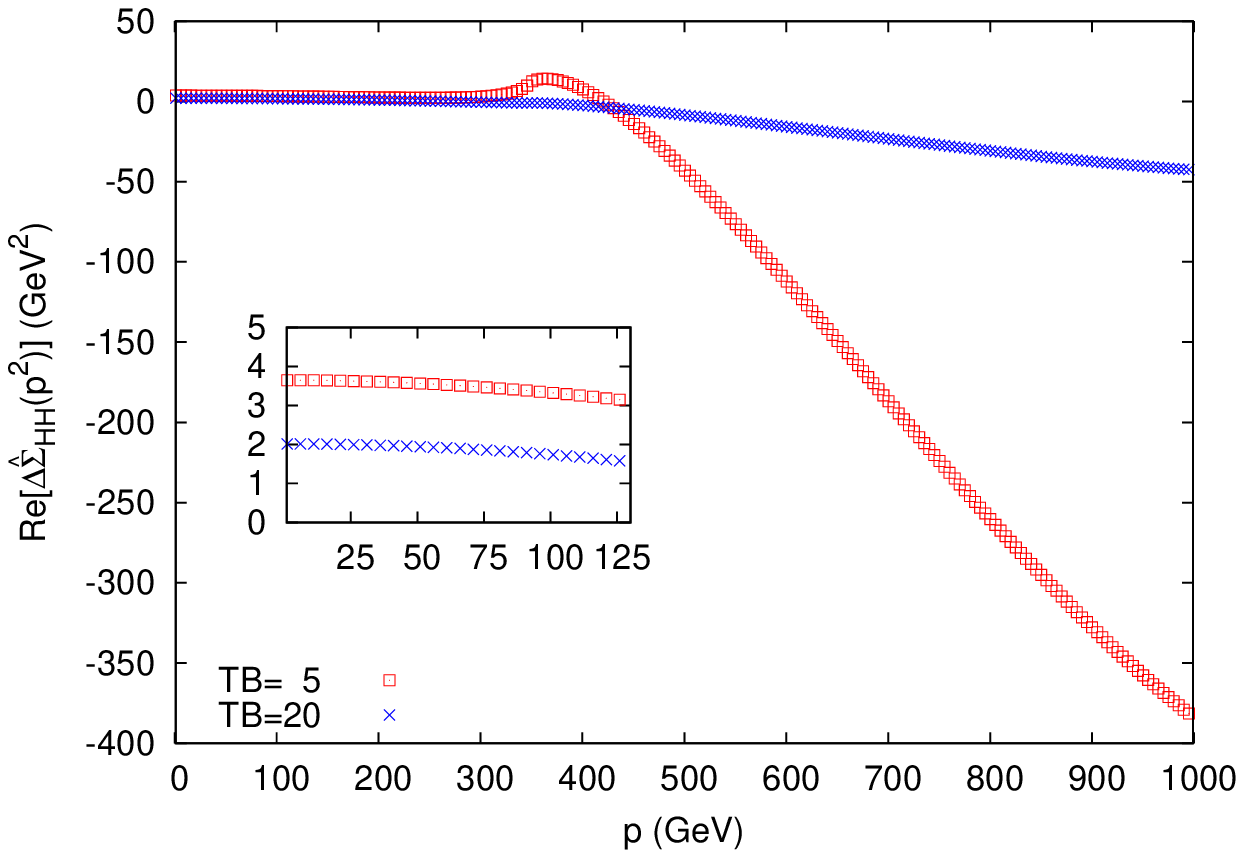}
\includegraphics[width=0.49\textwidth]{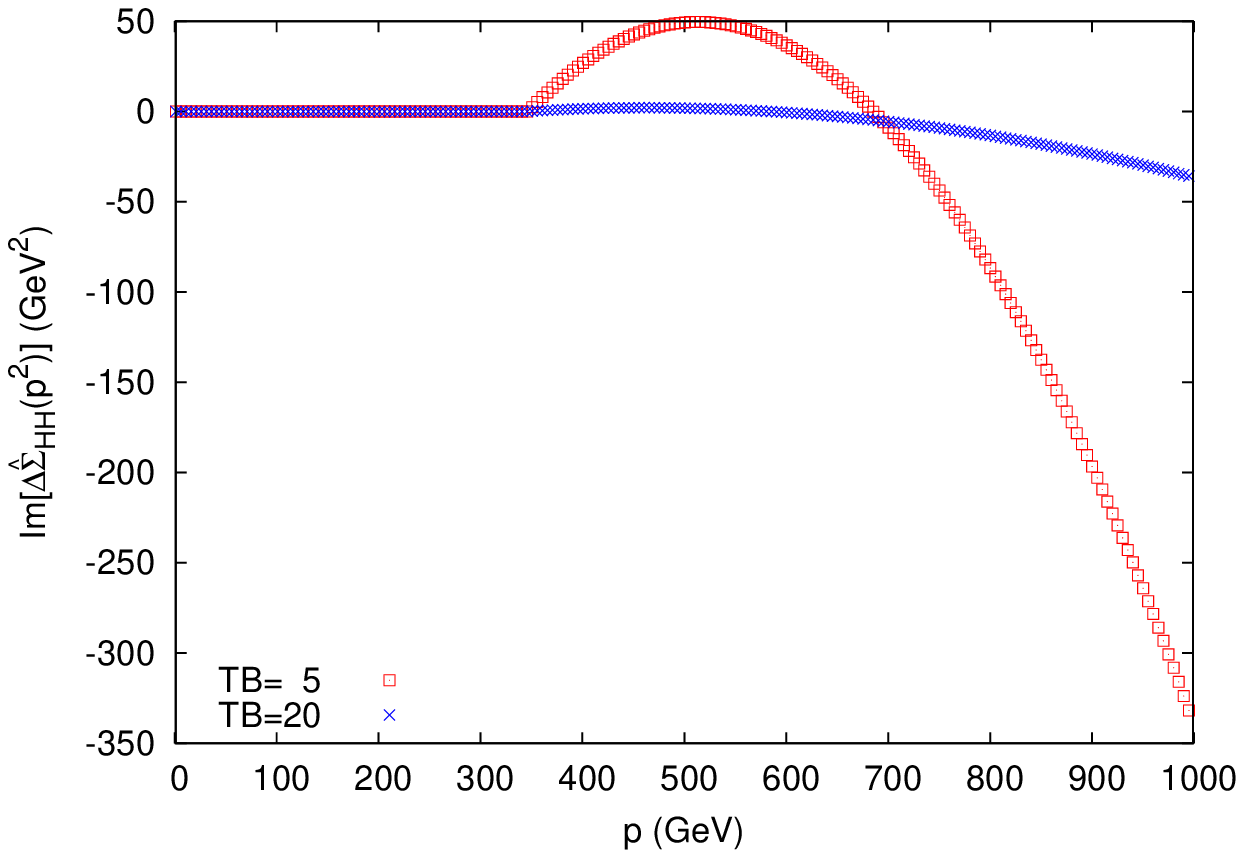}
\\[2em]
\caption{Momentum dependence of the real (left column) and 
imaginary (right column) parts of the two-loop selfenergies 
$\De\hat{\Sigma}_{hh},\De\hat{\Sigma}_{hH},\De\hat{\Sigma}_{HH}$, within
scenario 1,  
for $\tb=5$ (red squares) and $\tb = 20$ (blue crosses) and 
$\MA=250 \gev$. 
One can see that the selfenergies change 
substantially beyond the threshold at $p^2= (2\mt)^2$.}
\label{fig:se_scenario1reim}
\end{figure}

Scenario~1 is oriented at the \mhmax\ scenario described
in~\citere{Carena:2013qia}. 
We use the following parameters:\\[-2.5em]
\begin{align}
\mt &= 173.2\gev,\; \msusy=1\tev,\; \Xt =2\,\msusy\; , \non \\
\Mgl &= 1500\gev,\; \mu = 200\gev~, 
\end{align}
leading to stop mass values of
\begin{align}
\mste &= 826.8\gev,\;  \mstz= 1173.2\gev\, .\non
\end{align}
With the introduction of the momentum 
dependence, thresholds occur in the self-energy diagrams 
when the external momentum $p=\sqrt{p^2}$,
in the time-like region, 
is such that a cut of the diagram would correspond to 
on-shell production of the massive particles of the cut propagators.
The resulting imaginary parts  enter in the search for 
the complex poles of the inverse 
propagator matrix of the Higgs bosons.
Therefore it is interesting to study the 
behaviour of the real and imaginary parts of the self-energies.
In Fig.~\ref{fig:se_scenario1reim} we show the momentum dependent 
parts of the renormalized two-loop self-energies in the physical
basis, \refeq{eq:DeltaSE} 
for two different values of $\tb$, $\tan\beta=5$ and $\tan\beta=20$, at 
a fixed  $A$-boson mass $\MA=250$ GeV.
The data points are not connected by a line in order to show
that each numerical point is obtained from a calculation
of the 34 analytically unknown integrals with the program \sd{}. 
The inlays in Fig.~\ref{fig:se_scenario1reim} magnify the region
$p^2\leq (125 \gev)^2$, where one can observe that for $p^2\to 0$, 
the subtracted self-energies are not exactly 
zero.
As mentioned in subsection \ref{sec:fh}, this is due to the fact that the 
on-shell renormalization condition for the $A$-boson self-energy 
is defined differently with regard to the calculation without momentum
dependence.  
The resulting constant contributions are additionally suppressed by factors 
$\SQb$, $\Sbe\Cb$ and $\CQb$ appearing in the counter terms 
$\dmesqt$, $\dmezsqt$ and $\dmzsqt$, respectively, 
according to \refeqs{masscounterterms} in the Appendix.

\smallskip
The imaginary part is independent of
the $A$-boson mass, as this mass parameter 
solely appears in the counterterms of 
\DRbar\ renormalized quantities and 
the $\dMAsqt$ counterterm, where
only the real part contributes. 
Therefore, the imaginary parts displayed in Fig.~\ref{fig:se_scenario1reim}
do not contain additional constant terms.
As to be expected, the imaginary parts are zero below the 
$t\bar{t}$ production threshold at $p=2\,m_t$, 
which results from the fact that the top mass is the 
smallest mass appearing in the loops. 
Beyond this threshold,
the imaginary parts are growing substantially with increasing $p^2$.
{}From these observations, the mass shifts in the region 
below the first threshold at $p=2\,m_t$ are expected not to be large.

\begin{figure}[ht!]
\centering
\includegraphics[width=0.6\textwidth]{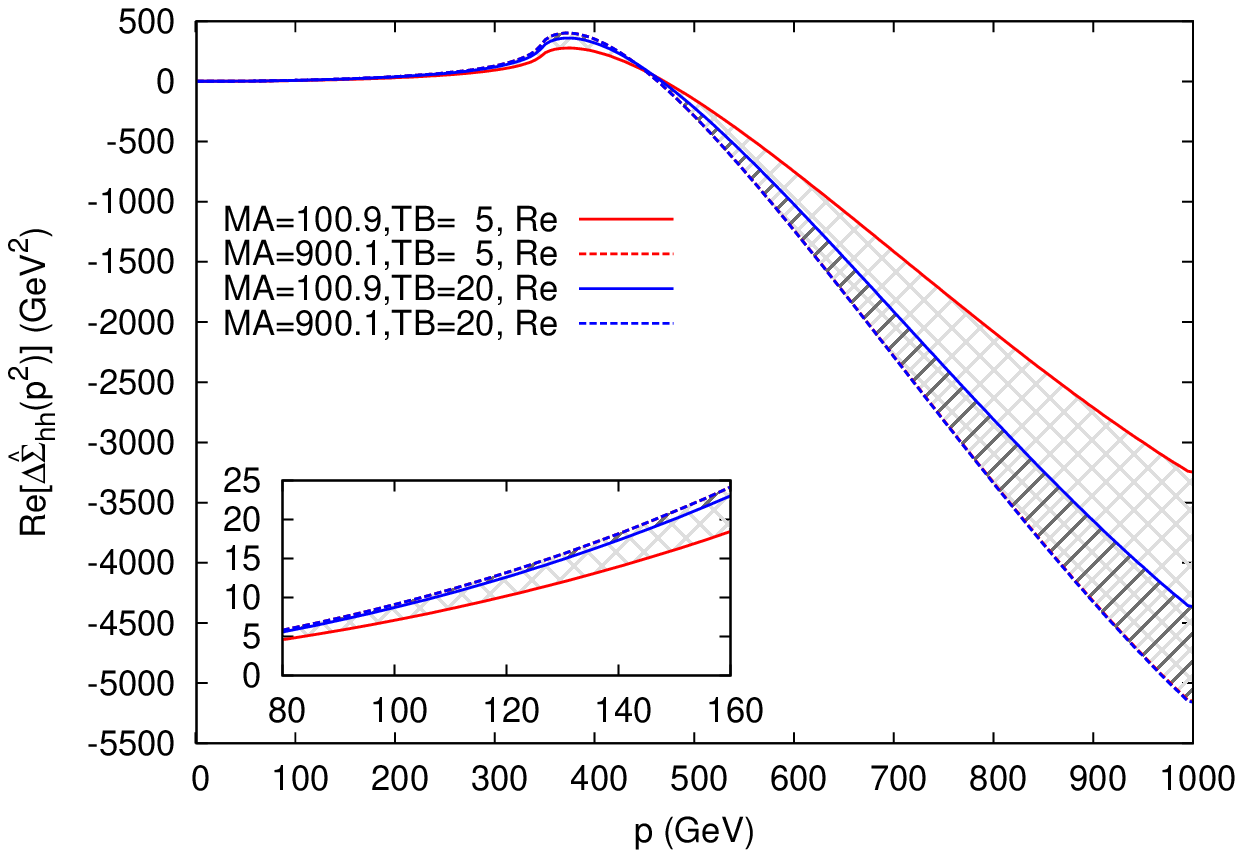}\\
\includegraphics[width=0.6\textwidth]{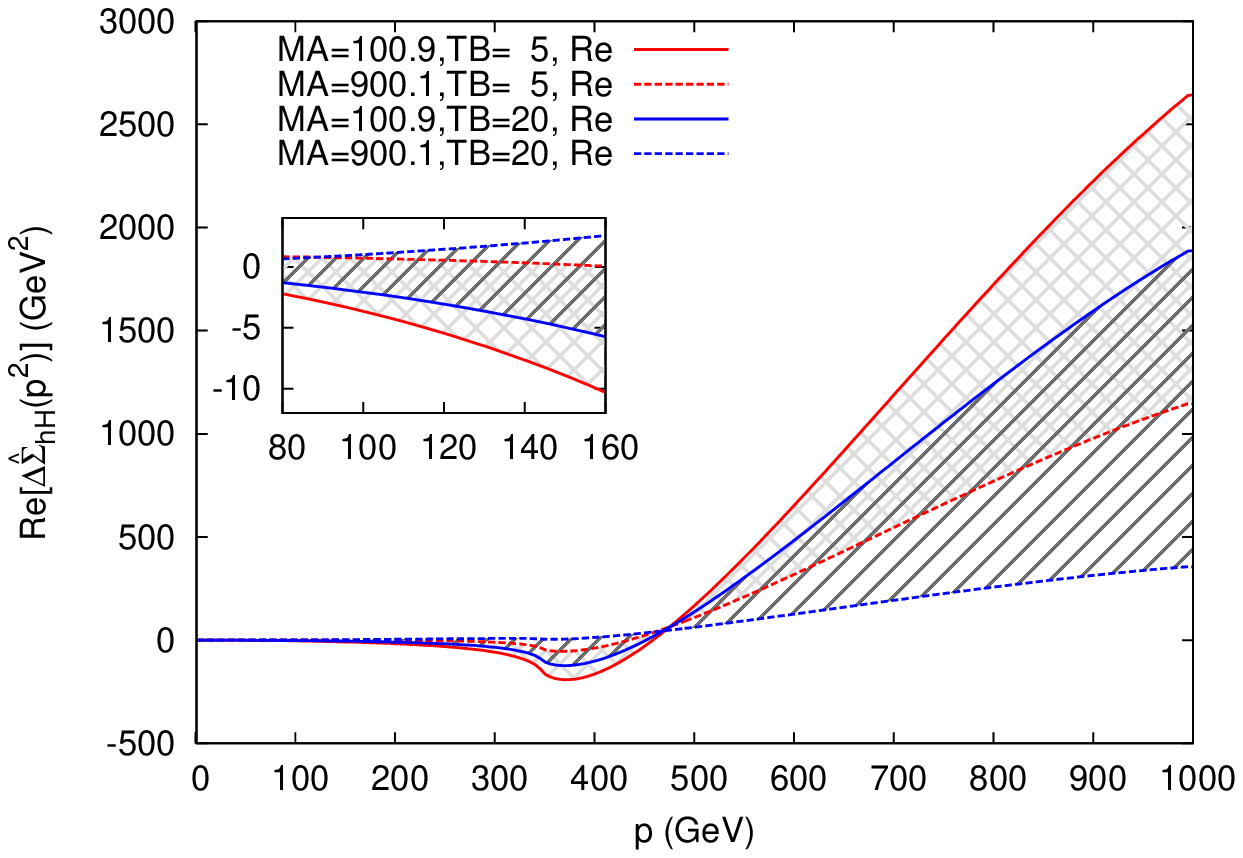}\\
\includegraphics[width=0.6\textwidth]{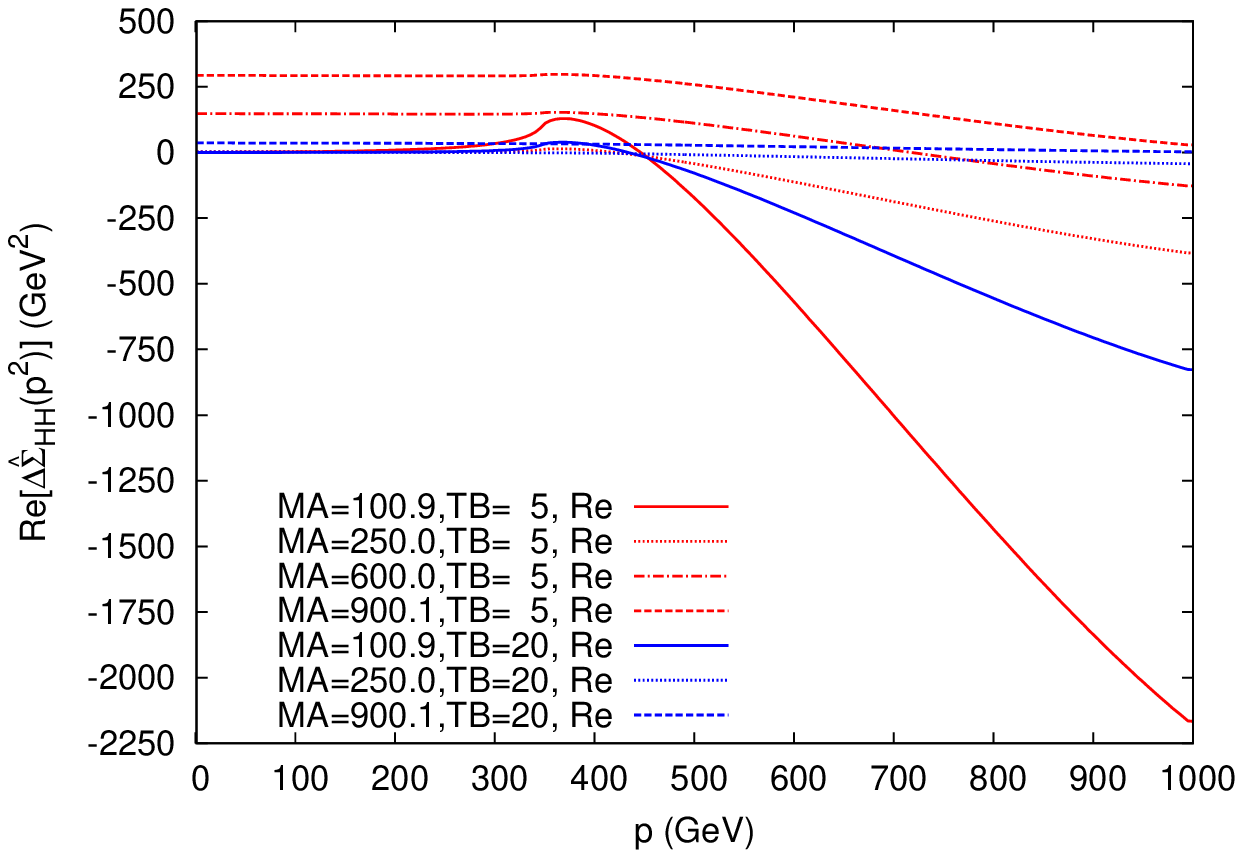}
\caption{Momentum dependence of the real part of the two-loop selfenergies 
$\De\hat{\Sigma}_{hh},\De\hat{\Sigma}_{hH},\De\hat{\Sigma}_{HH}$, within
scenario~1, 
for two different  values of $\tb$ and a range of $\MA$ values.
}
\label{fig:se_scenario1_madep}
\end{figure}

\smallskip
Similar results, now including a variation of $\MA$ are shown 
 in~\reffi{fig:se_scenario1_madep}. 
In the upper plot for $\De\ser{hh}$ and in the middle plot for $\De\ser{hH}$
the solid lines depict $\MA \sim 100 \gev$, 
while the dashed lines are for $\MA \sim 900 \gev$.
In these plots the light shading covers the range for $\tb = 5$, while
the dark shading for $\tb = 20$.
In the lower plot for $\De\ser{HH}$ we show results for 
$\MA \sim 100, 250, 600, 900 \gev$ as solid, dotted, dot-dashed, dashed
lines, respectively (and shading has been omitted).
For $\De\ser{hh}$ at low~$p$ values only a small variation with $\MA$ can
be observed. For $p$ and $\MA$ large, the 
contributions to the self-energy are bigger. In $\De\ser{hH}$
larger effects are observed at smaller $\MA$  for both small and 
large~$p$ values. For $\De\ser{HH}$, on the other hand, at low~$p$ values,
large effects can be observed for large $\MA$ due to the aforementioned
counterterm contribution $\sim \dMAsqt = \re\se{AA}^{(2)}(\MA^2)$. At
large~$p$, as before, small $\MA$ values give a more sizable contribution.

\begin{figure}[ht!]
\centering
\includegraphics[width=0.7\textwidth]{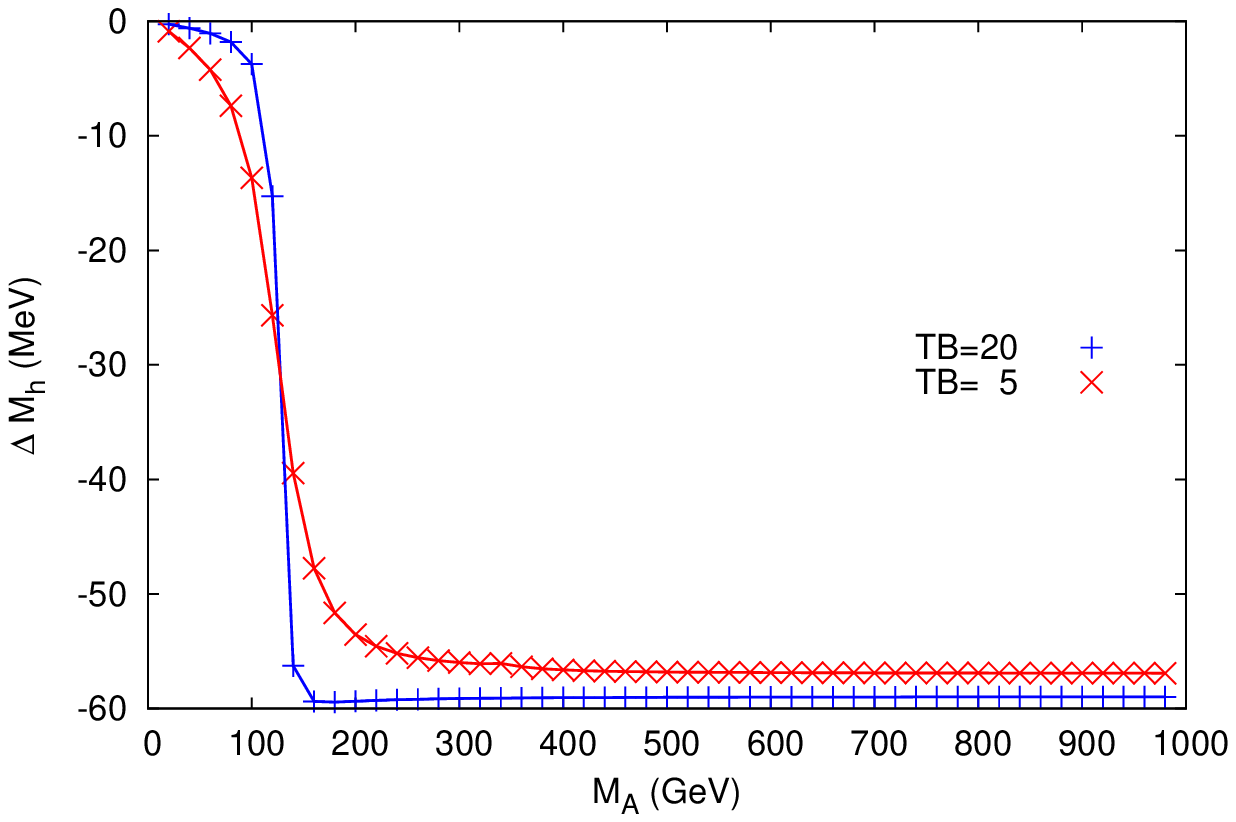}\\
\includegraphics[width=0.7\textwidth]{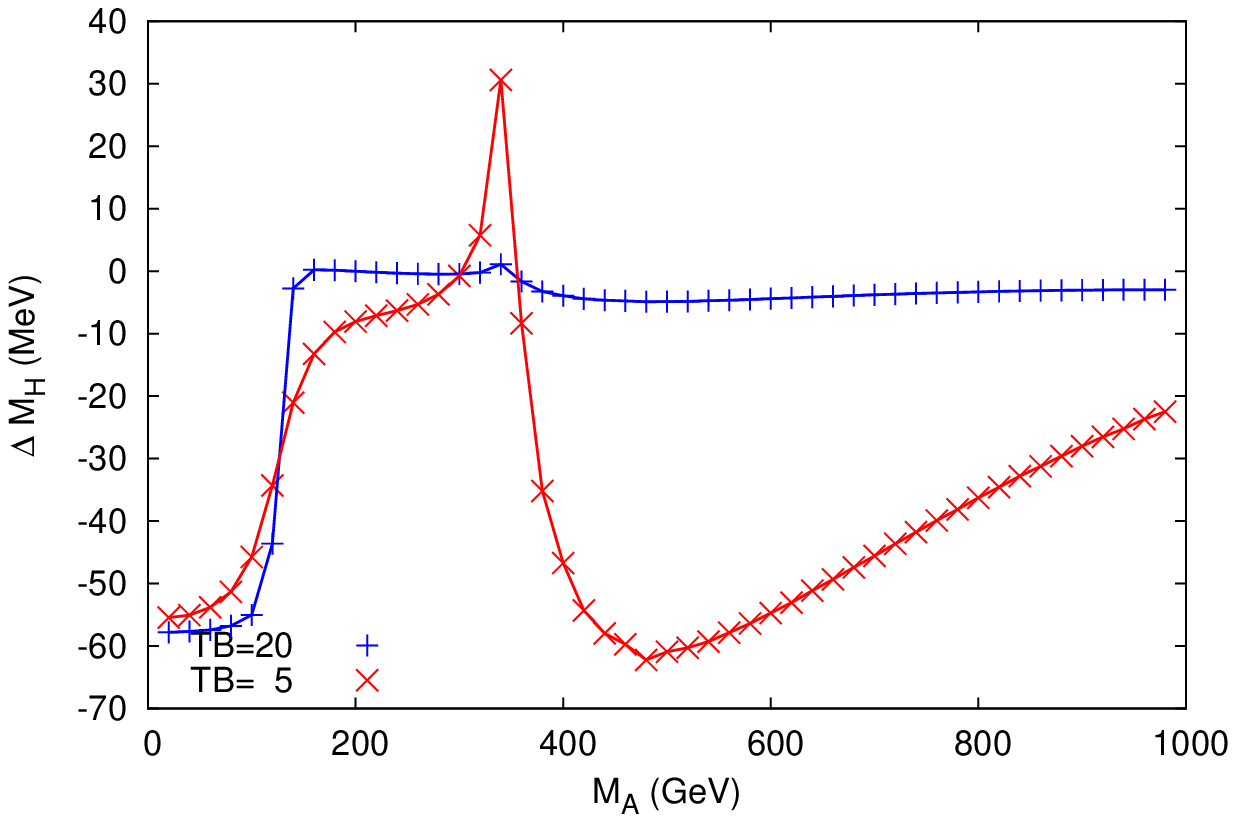}
\caption{Variation of the mass shifts $\Delta\Mh,\Delta\MH$ with the
  $A$-boson mass $\MA$ within scenario~1, 
for $\tb=5$ (red) and $\tb = 20$ (blue). The peak in $\Delta\MH$
originates from a threshold at $2\,\mt$.} 
\label{fig:shiftswithma}
\end{figure}

\smallskip
We now turn to the effects of our newly computed momentum-dependent
two-loop corrections on the Higgs-boson masses $M_{h,H}$ via the mass shifts
$\De\Mh$ and $\De\MH$. 
In \reffi{fig:shiftswithma} we
show $\De\Mh$ (upper plot) and $\De\MH$ (lower plot) as a function of
$\MA$ for $\tb = 5$ (blue) and $\tb = 20$ (red). 
In the \mhmax\ scenario for $\MA \gsim 200 \gev$ we find 
$\Delta\Mh \sim - 60 \mev$, i.e.\ of the size of the future experimental
precision, see \refeq{dMhILC}. The contribution to the heavy $\cp$-even
Higgs-boson is suppressed with $\tb$. 
While the size of $\De\MH$ becomes negligible for $\MA \gsim 150 \gev$
for $\tb = 20$, its variation is more pronounced for $\tb = 5$.
$\De\MH$ can
reach about $-60 \mev$ for very small or intermediate values of $\MA$
and steadily decreases for $\MA \gsim 500 \gev$.
The peak in $\De\MH$ for $\tb = 5$ originates from a threshold at
$2\,\mt$.

\begin{figure}[htb!]
\centering
\includegraphics[width=0.7\textwidth]{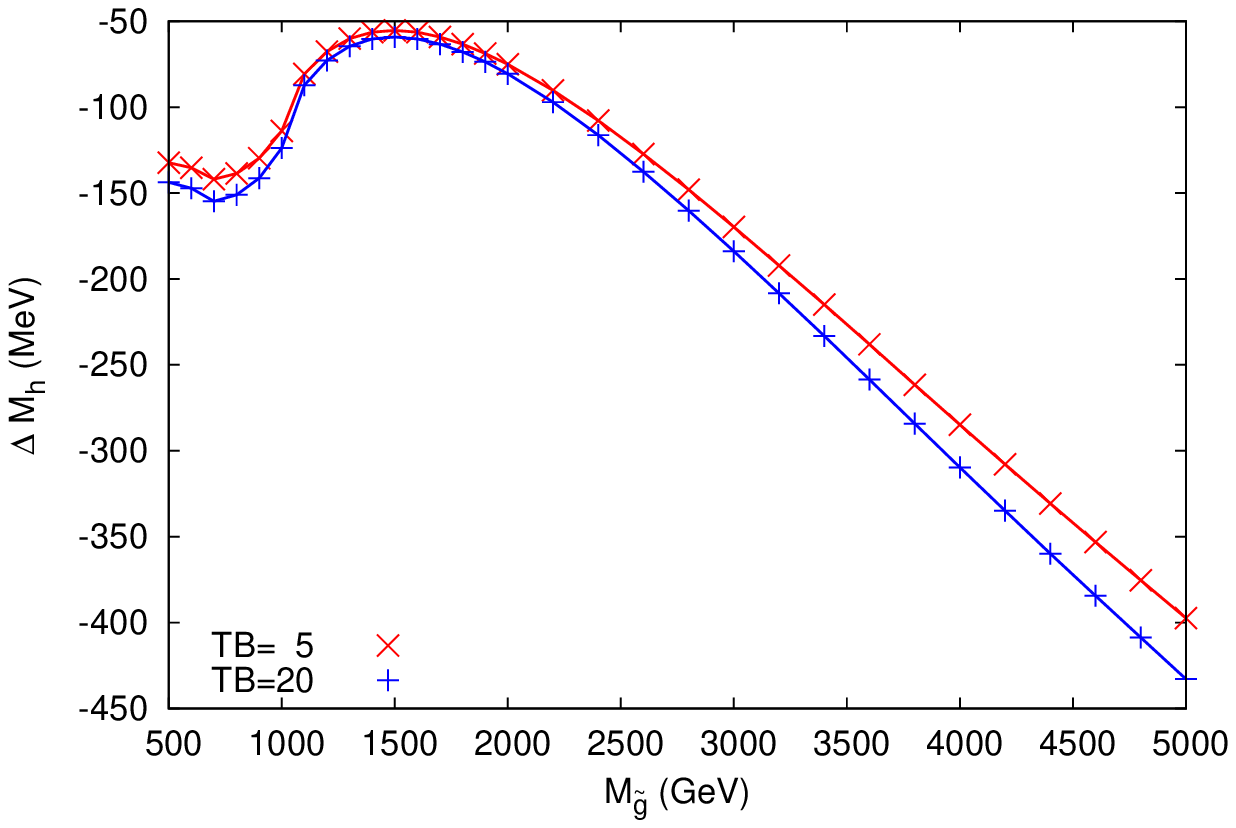}
\includegraphics[width=0.7\textwidth]{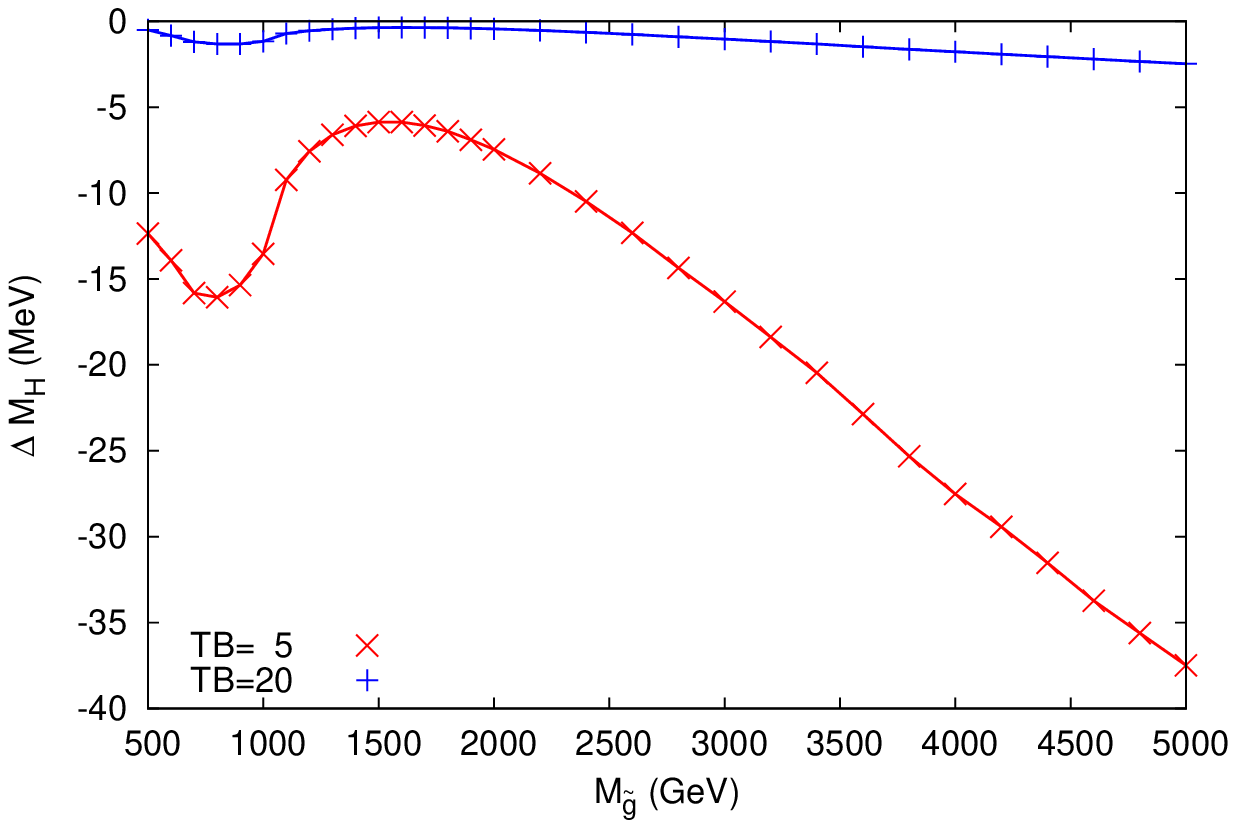}
\caption{Variation of the mass shifts $\Delta\Mh,\Delta\MH$ with the
  gluino mass, within scenario 1, for two different  values of
  $\tb=5,20$ and $\MA = 250 \gev$.
}
\label{fig:variationmgluino}
\end{figure}

\medskip
Finally, in scenario~1, we analyze the dependence of $\Mh$ and
$\MH$ on the gluino mass, $\Mgl$. The results are shown in
\reffi{fig:variationmgluino} for $\De\Mh$ (upper plot) and $\De\MH$
(lower plot) for $\MA = 250 \gev$,
with the same color coding as in \reffi{fig:shiftswithma}.
In the upper
plot one can observe that the effects are particularly small for the
default value of $\Mgl$ in scenario~1.
More sizeable shifts occur for larger gluino masses,   
by more than $-400 \mev$ for $\Mgl \gsim 4 \tev$, 
reaching thus the level of
the current experimental accuracy in the Higgs-boson mass 
determination.
The corrections to $\MH$, for the given value of $\MA = 250 \gev$
do not exceed $-50 \mev$ in the considered $\Mgl$ range.

\clearpage

\begin{figure}[htb!]
\centering
\includegraphics[width=0.9\textwidth]{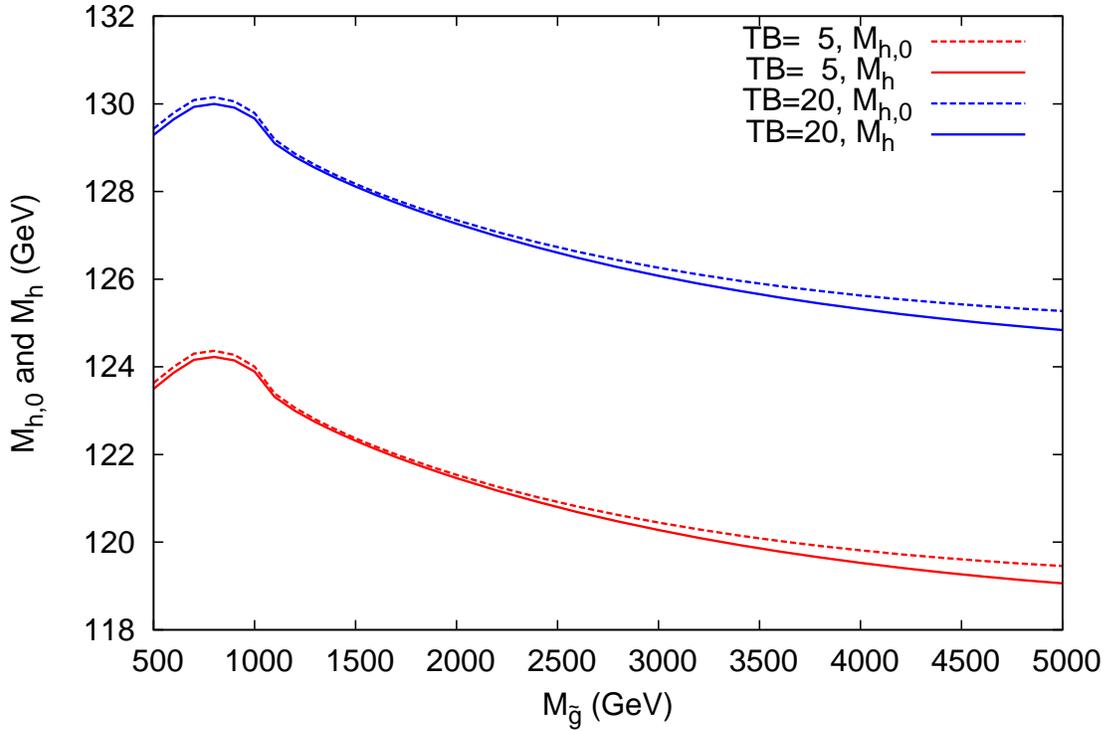}
\caption{
Variation of $\Mh$ and $M_{h,0}$ as a function of $\Mgl$ within
scenario 1, for $\tb = 5, 20$ and $\MA = 250 \gev$. 
}
\label{fig:gluino-ana}
\end{figure}

The dependence of the light $\cp$-even Higgs-boson mass on $\Mgl$ is analyzed in
\reffi{fig:gluino-ana} for $\tb = 5, 20$ and $\MA = 250 \gev$. Here we
show as dashed lines the results for $M_{h,0}$ (i.e.\ without the newly
obtained momentum dependent two-loop corrections) and as solid lines the
results for $\Mh$ (i.e.\ including the new corrections). While a maximum
of the Higgs-boson mass can be observed around $\Mgl \sim 800 \gev$, in
agreement with the original definition of the \mhmax\
scenario~\cite{benchmark2}, a downward shift by more than $4 \gev$ is
found for $\Mgl \sim 5 \tev$. Such a strong effect is due to a
(squared) logarithmic dependence of the \order{\alt\als} corrections
evaluated at $p^2 = 0$, as given in Eq.~(73) of \citere{mhiggslong}. 
In \reffi{fig:gluino-ana} it can be seen that the size of the momentum
dependent two-loop corrections similarly grows with $\Mgl$, reaching
$\sim 400 \mev$, as was shown above in
\reffi{fig:variationmgluino}. Consequently, the logarithmic dependence
of the light $\cp$-even Higgs-boson mass on the gluino mass that
was found analytically for the 
\order{\alt\als} corrections at $p^2 = 0$, is now also found numerically
for the momentum dependent two-loop corrections.

\clearpage


\subsection{Scenario 2: light stops}

Scenario~2 is oriented at the ``light-stop scenario'' of
\citere{Carena:2013qia}%
\footnote{
While the original scenario in \citere{Carena:2013qia} is challenged by
recent scalar-top searches at ATLAS and CMS, a small modification in the
gaugino-mass parameters (which play no or only a very minor role here)
to $M_1 = 340 \gev$, $M_2 = \mu = 400 \gev$ 
leads to a SUSY spectrum that is very difficult to test at the LHC.
}%
.~We use the following parameters:
\begin{align}
\mt &= 173.2\gev,\; \msusy= 0.5 \tev,\; \Xt =2\,\msusy\; , \non \\
\Mgl &= 1600\gev,\; \mu = 200\gev~, 
\end{align}
leading to stop mass values of
\begin{align}
\mste &= 326.8 \gev,\;  \mstz= 673.2 \gev\, .\non
\end{align}

\begin{figure}[htb!]
\centering
\includegraphics[width=0.49\textwidth]{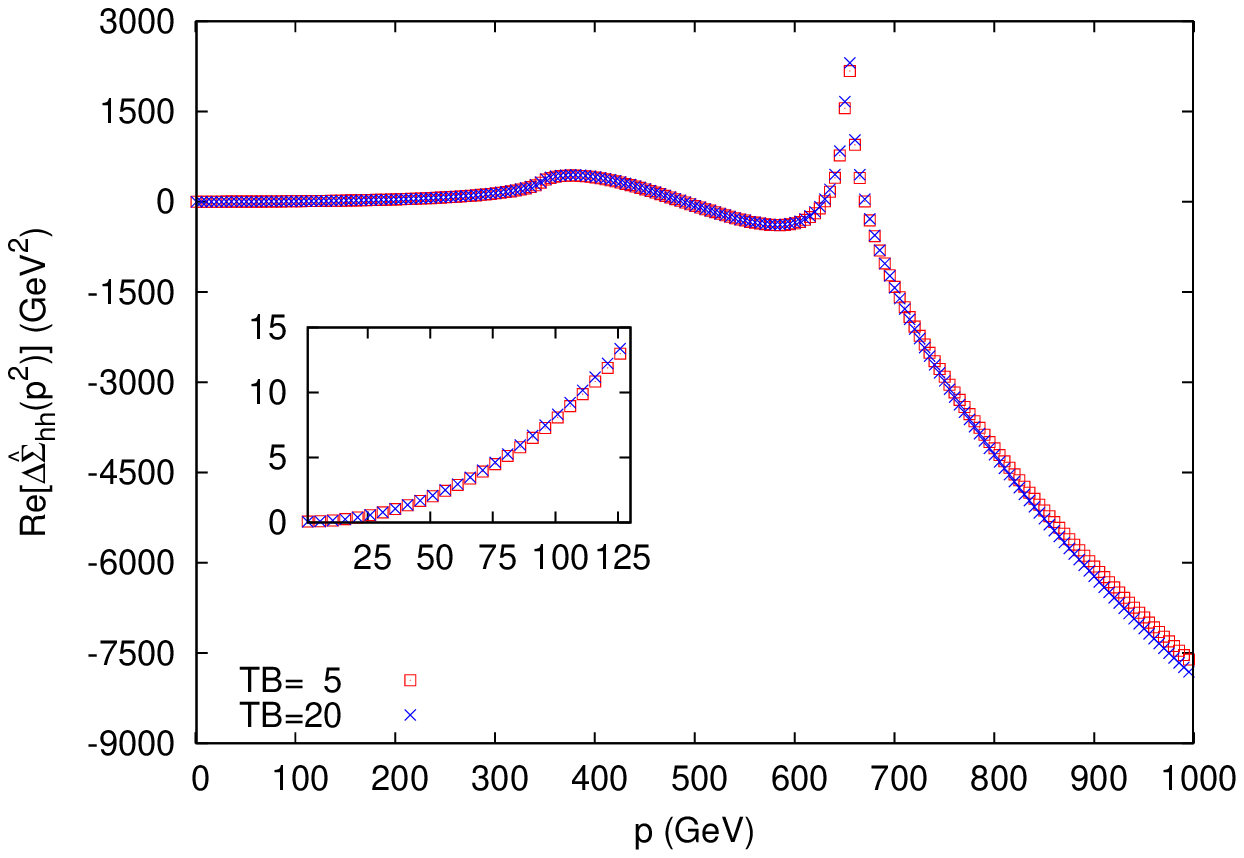}
\includegraphics[width=0.49\textwidth]{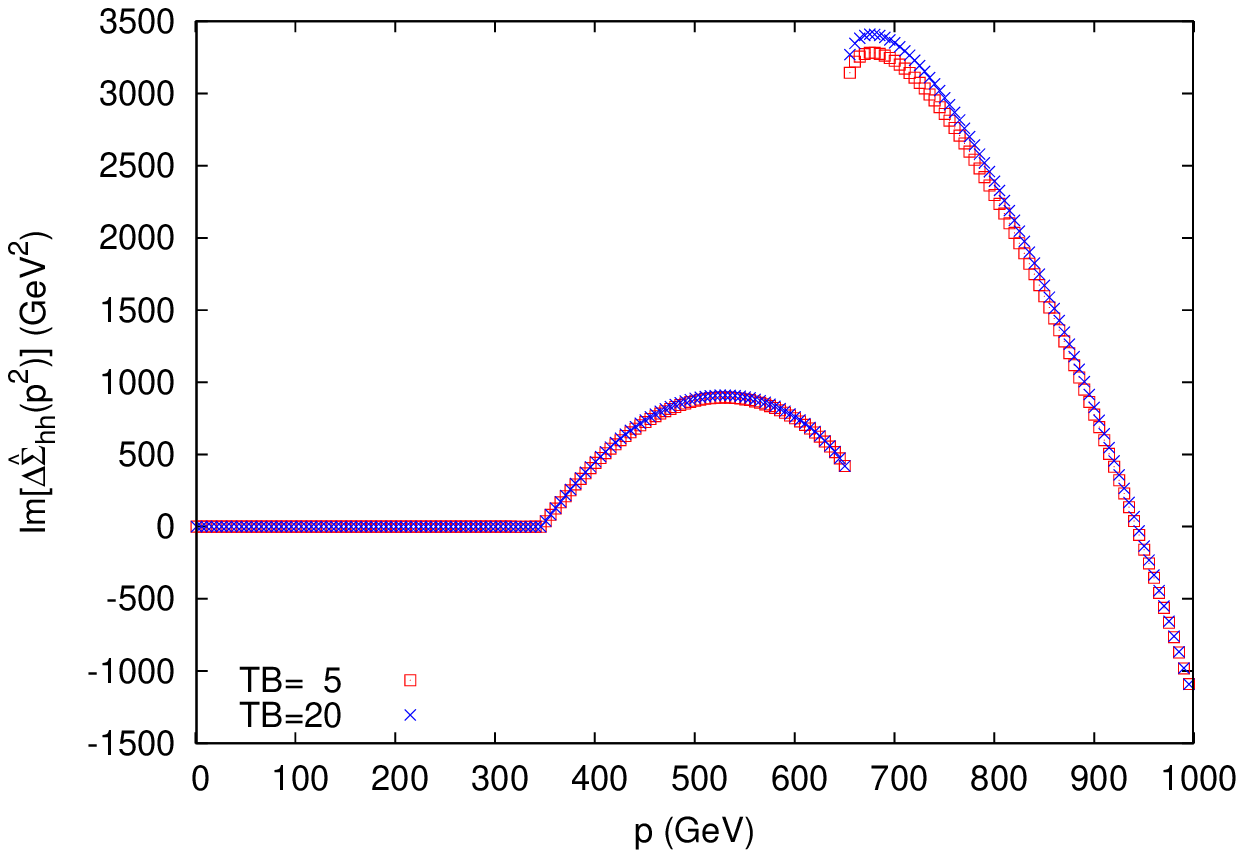}
\\
\includegraphics[width=0.49\textwidth]{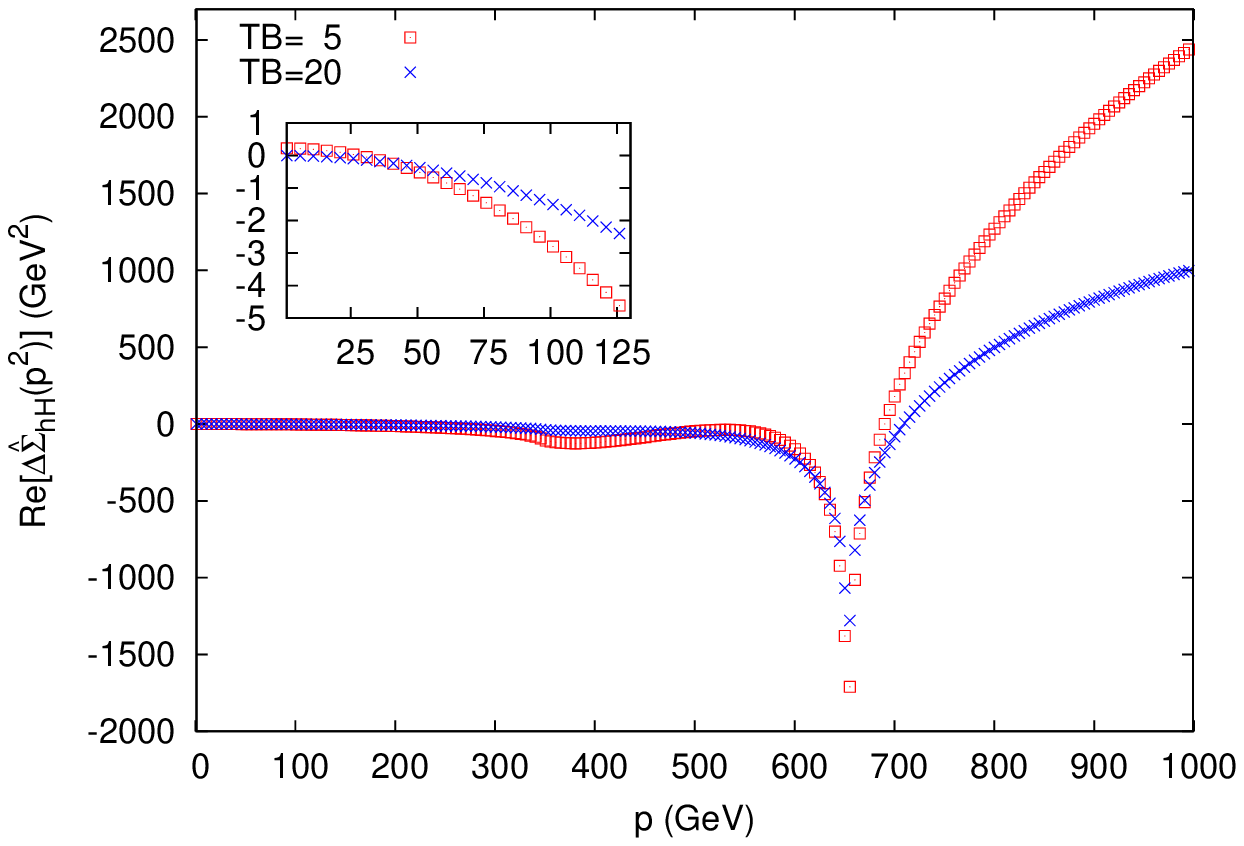}
\includegraphics[width=0.49\textwidth]{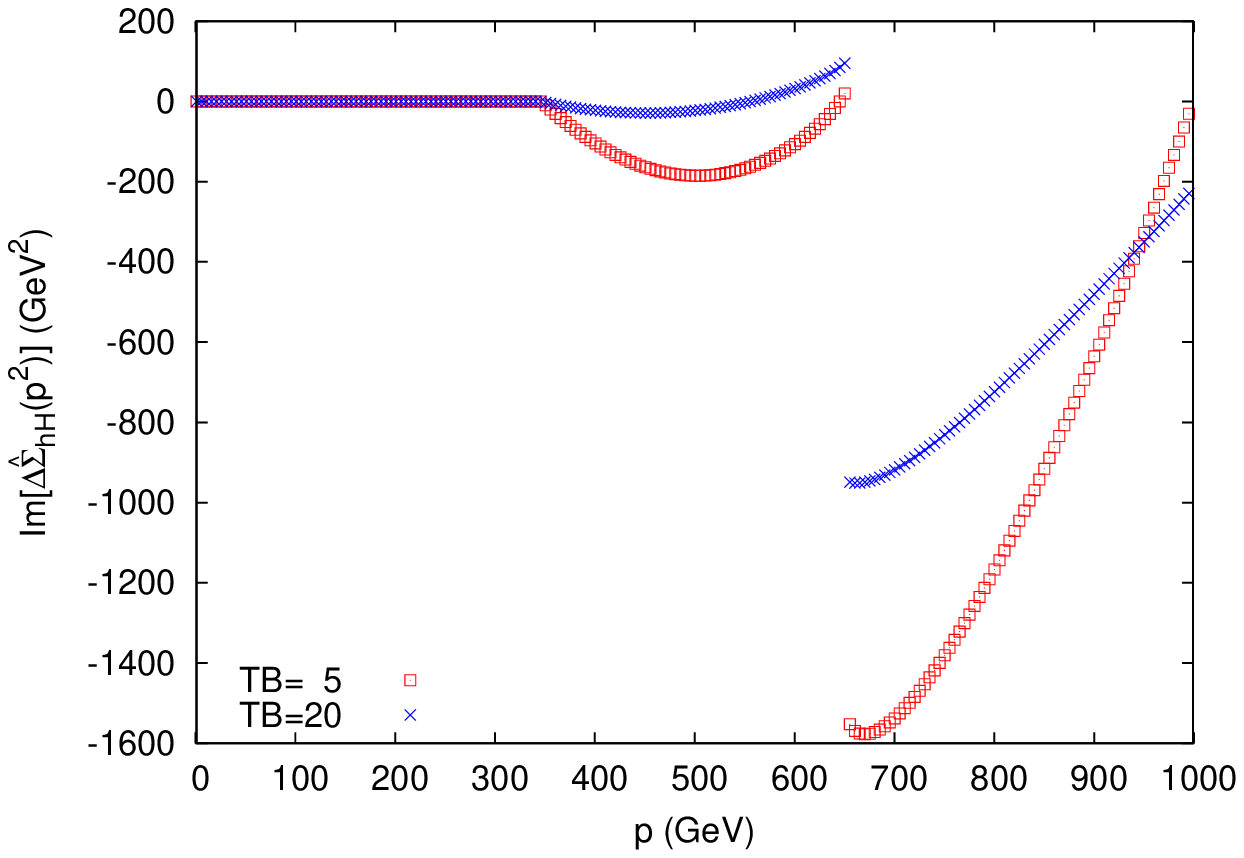}
\\
\includegraphics[width=0.49\textwidth]{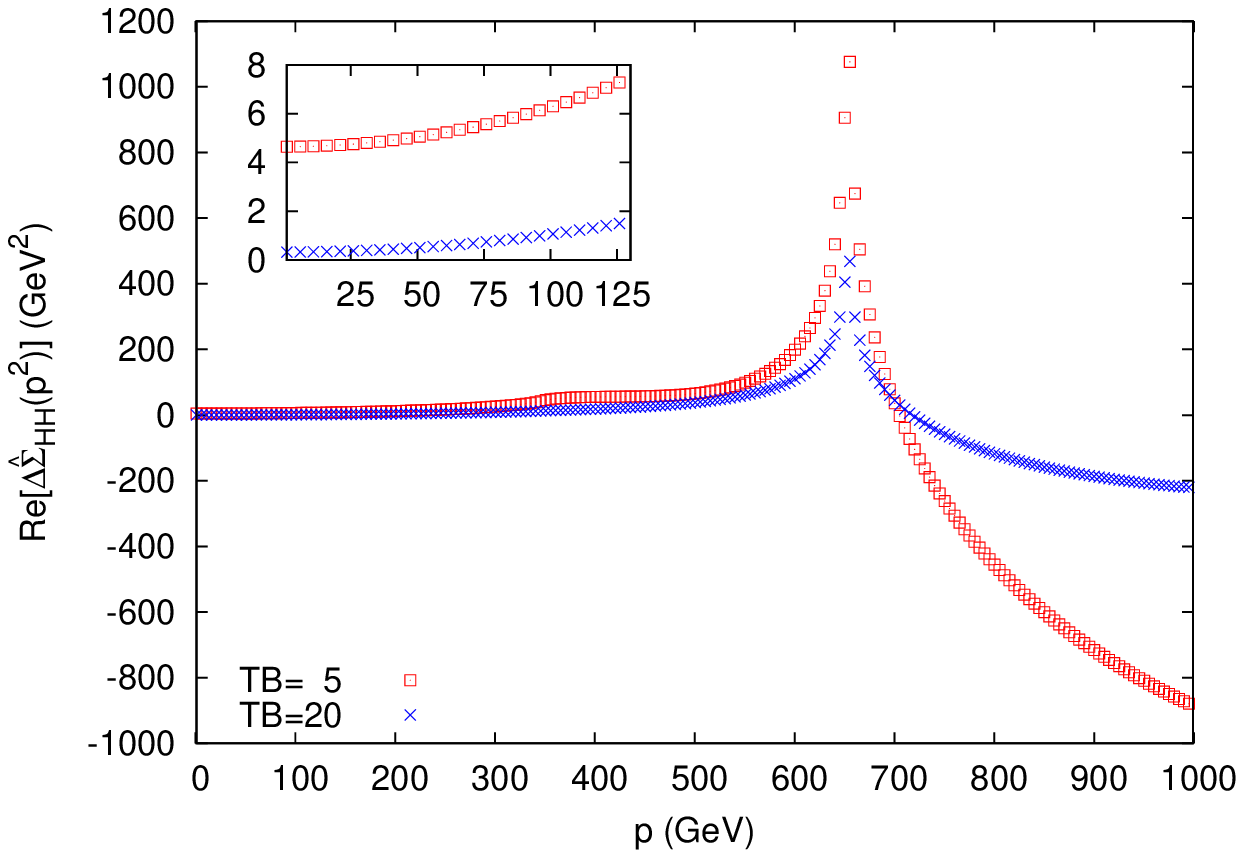}
\includegraphics[width=0.49\textwidth]{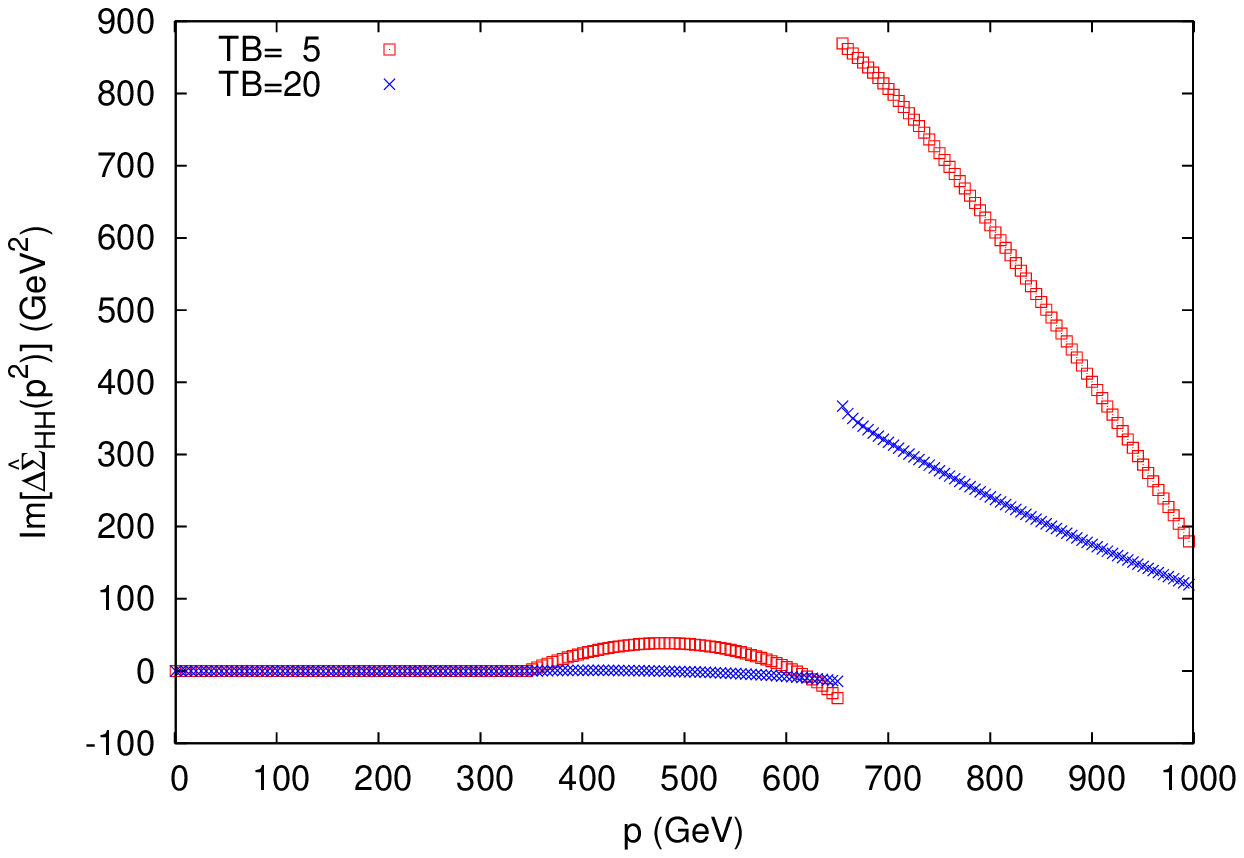}
\\
\caption{Momentum dependence of the real and imaginary parts of the
  two-loop self-energies 
$\De\hat{\Sigma}_{hh},\De\hat{\Sigma}_{hH},\De\hat{\Sigma}_{HH}$ within 
scenario 2, with $\tb=5,20$ and $\MA=250 \gev$ with the same color
coding as in \reffi{fig:se_scenario1reim}.}
\label{fig:se_scenario2reim}
\end{figure}

Scenario~2 is analyzed with the same set of plots shown for scenario~1
in the previous subsection.
The effects of the new momentum dependent two-loop contributions on the
renormalized Higgs-boson self-energies, $\De\ser{ab}(p^2)$, are shown in
\reffi{fig:se_scenario2reim}. As before, we show the results separately 
for the real and imaginary parts of the self-energies. 
An additional threshold beyond the top-mass threshold appears at 
$p = 2\, \mste$, where the discontinuity stems from the derivative
of the imaginary part of the $B_0$ function(s).
Analogously to scenario~1, the largest contributions in the 
region below $200\gev$ arise in the real part of 
$\De\hat{\Sigma}_{hh}$ amounting to about  $15\gev^2$ at $p=125\gev$, 
where the dependence on the value of $\tanb$ is rather weak.

\begin{figure}[htb!]
\centering
\includegraphics[width=0.6\textwidth]{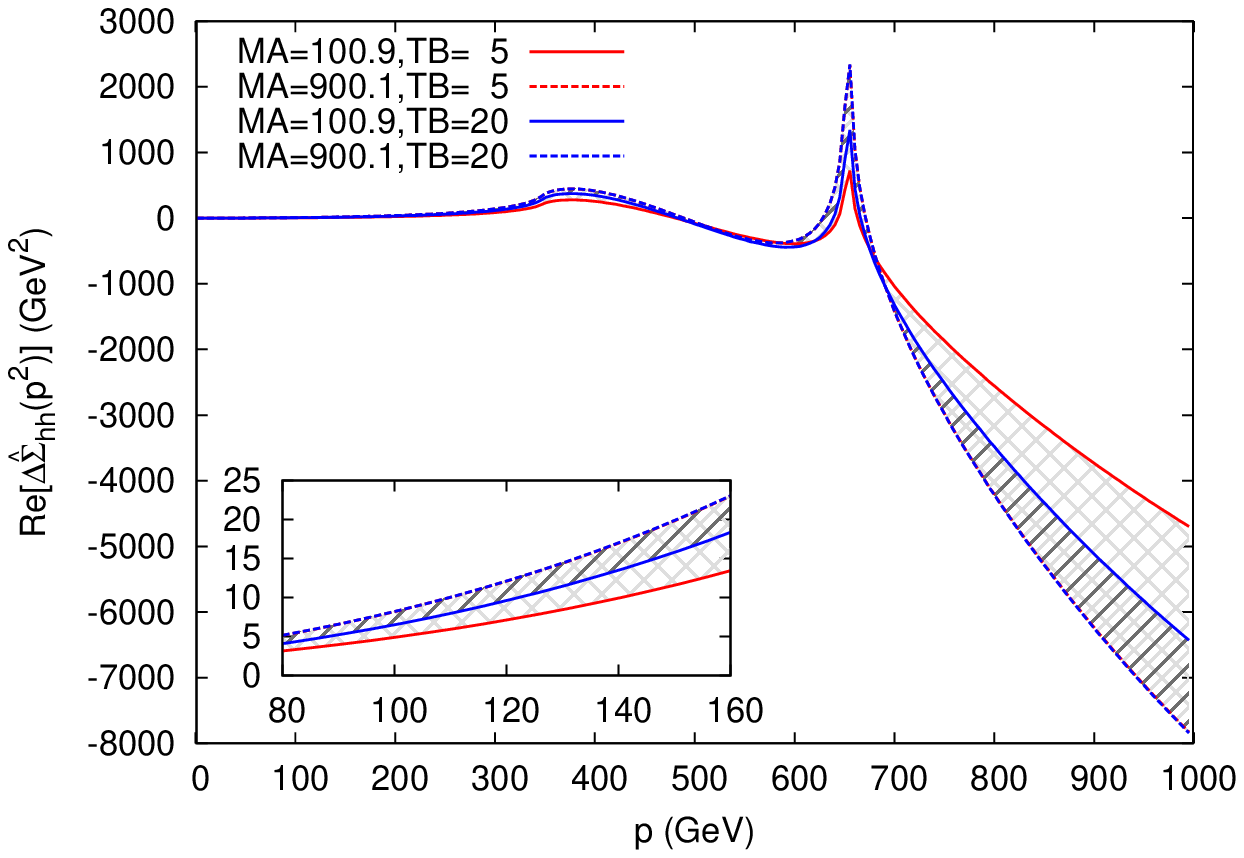}\\
\includegraphics[width=0.6\textwidth]{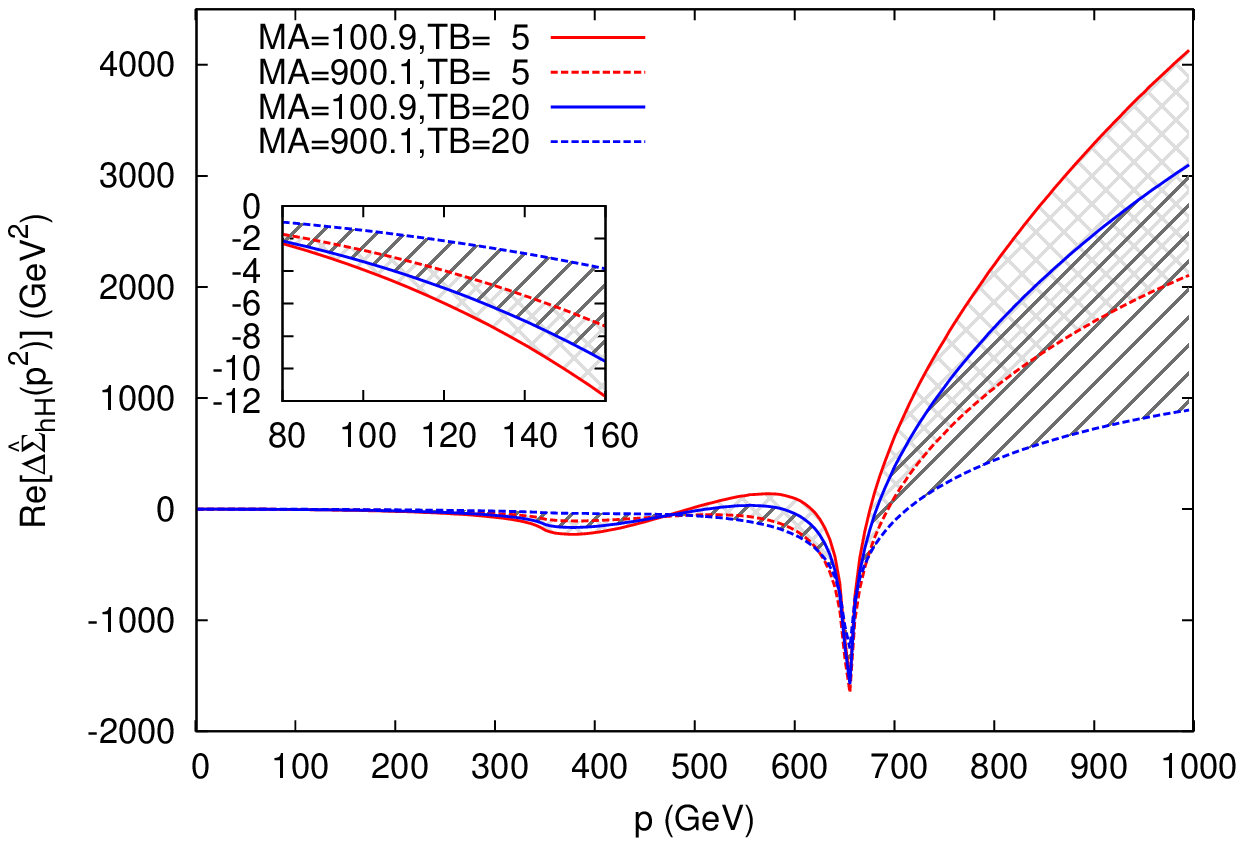}\\
\includegraphics[width=0.6\textwidth]{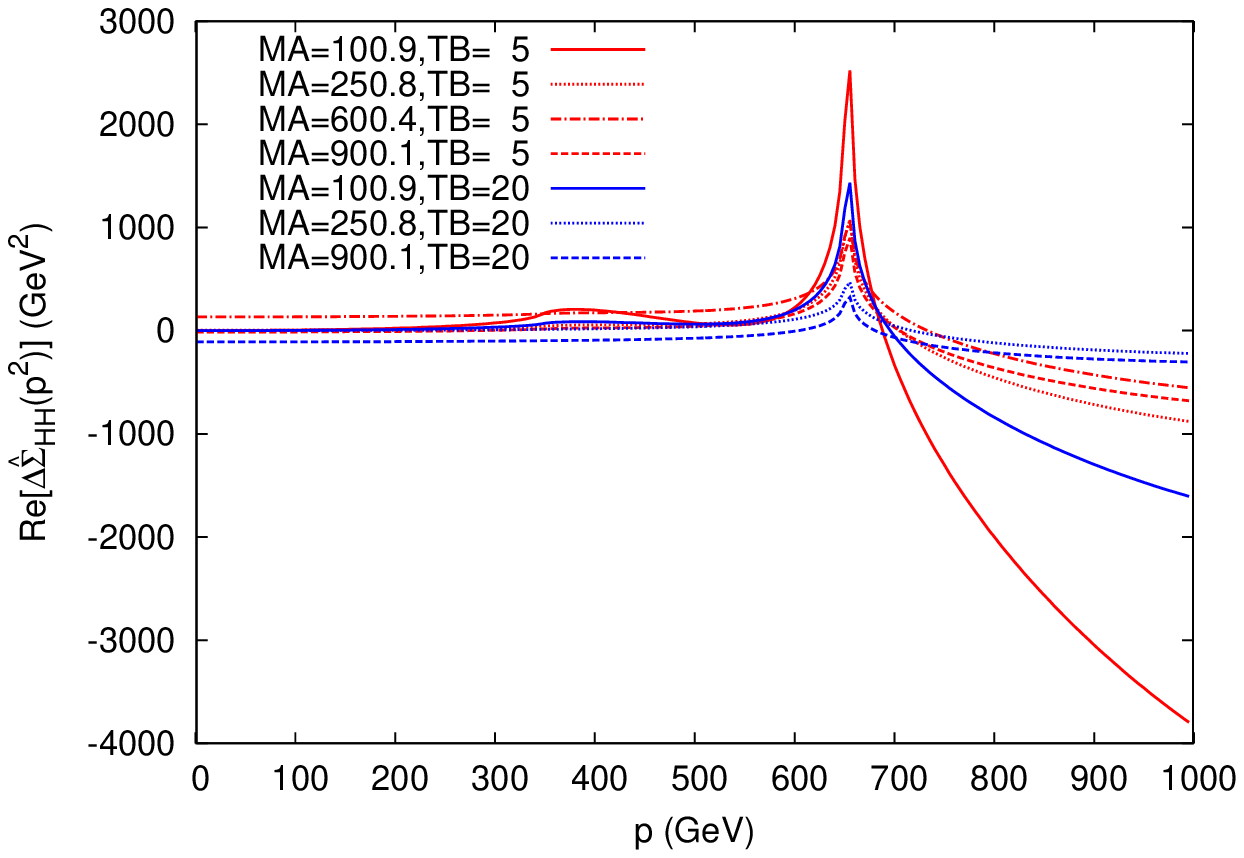}
\caption{Momentum dependence of the real parts of the two-loop self-energies 
$\De\hat{\Sigma}_{hh},\De\hat{\Sigma}_{hH},\De\hat{\Sigma}_{HH}$ in scenario~2 
for two different  values of $\tb$ and various values of $\MA$ (see text).}
\label{fig:se_scenario2_madep}
\end{figure}

\smallskip
The dependence of $\De\ser{ab}(p^2)$ on $\MA$ is shown in
\reffi{fig:se_scenario2_madep}, using the same line styles as in
\reffi{fig:se_scenario1_madep}. The curves show the same qualitative 
behaviour as in \reffi{fig:se_scenario2reim}, exhibiting again  
the new threshold at $p = 2\, \mste$. 
In general, outside the threshold region
the effects in scenario~2 are slightly smaller than in scenario~1.

\begin{figure}[ht!]
\centering
\includegraphics[width=0.7\textwidth]{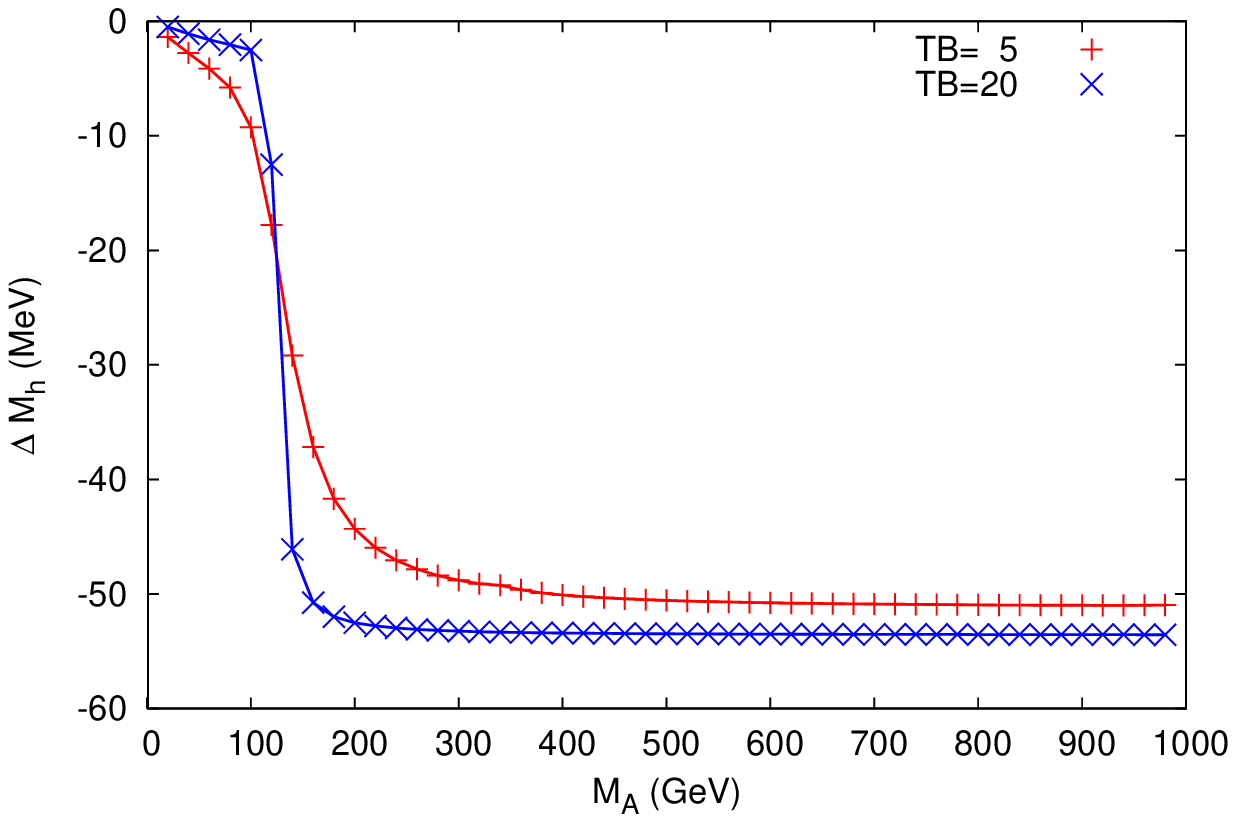}\\
\includegraphics[width=0.7\textwidth]{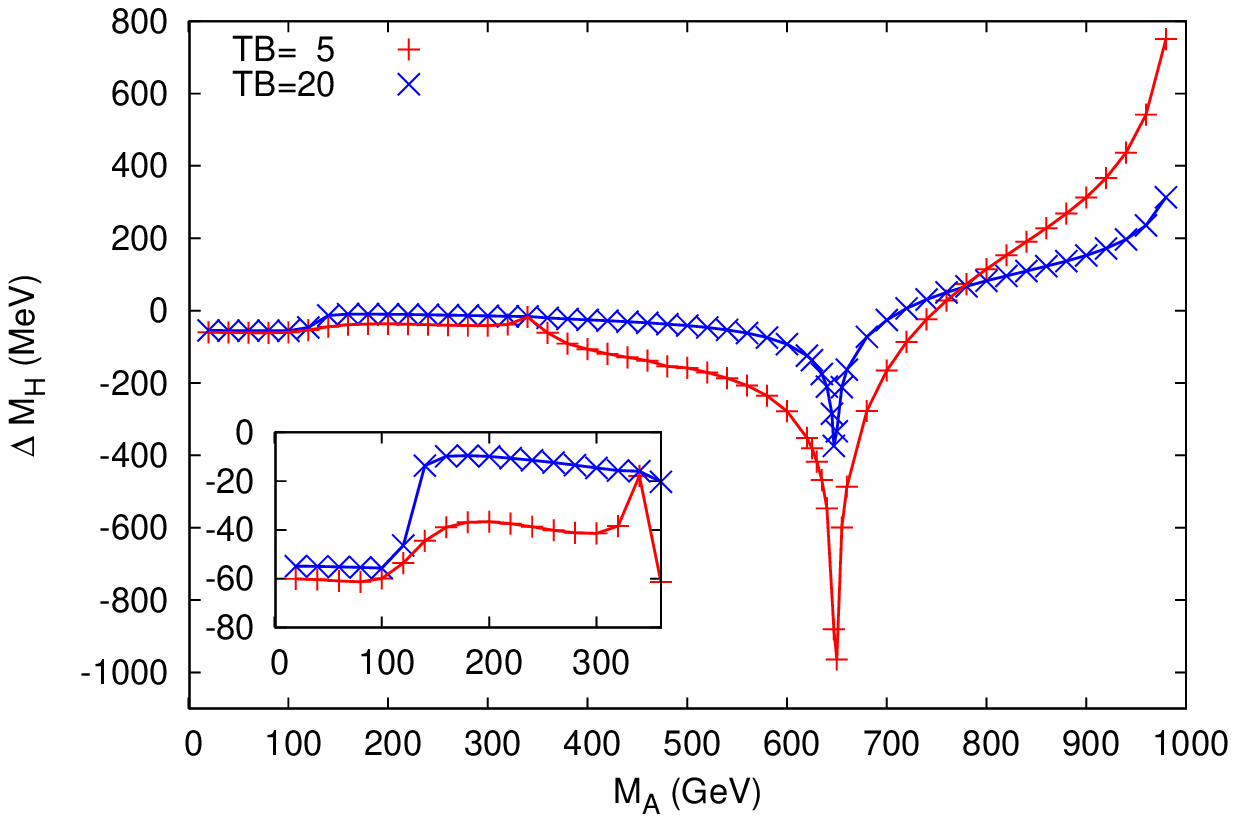}
\caption{Variation of the mass shifts $\Delta\Mh,\Delta\MH$ with  the $A$-boson mass $\MA$ within scenario~2,
for two different  values of $\tb=5,20$. 
}
\label{fig:scen2shiftswithma}
\end{figure}

\bigskip
We now turn to the effects on the physical neutral $\cp$-even Higgs
boson masses. In \reffi{fig:scen2shiftswithma} we show the results for
$\De\Mh$ (upper plot) and $\De\MH$ (lower plot) as a function of $\MA$
(with the same line styles as in \reffi{fig:shiftswithma}). As can
be expected from the previous figures, the effects on $\Mh$ and $\MH$
are in general slightly smaller in scenario~2 than in scenario~1, where
$\De\Mh$ still reaches the anticipated ILC accuracy, see
\refeq{dMhILC}. 
For $\De\MH$ around the threshold $p = 2\,\mste$ the largest shift
  of $\sim - 1 \gev$ can be observed. However, this shift is still below
  the anticipated mass resolution at the LHC~\cite{cmsHiggs}. 

\begin{figure}[htb!]
\centering
\includegraphics[width=0.7\textwidth]{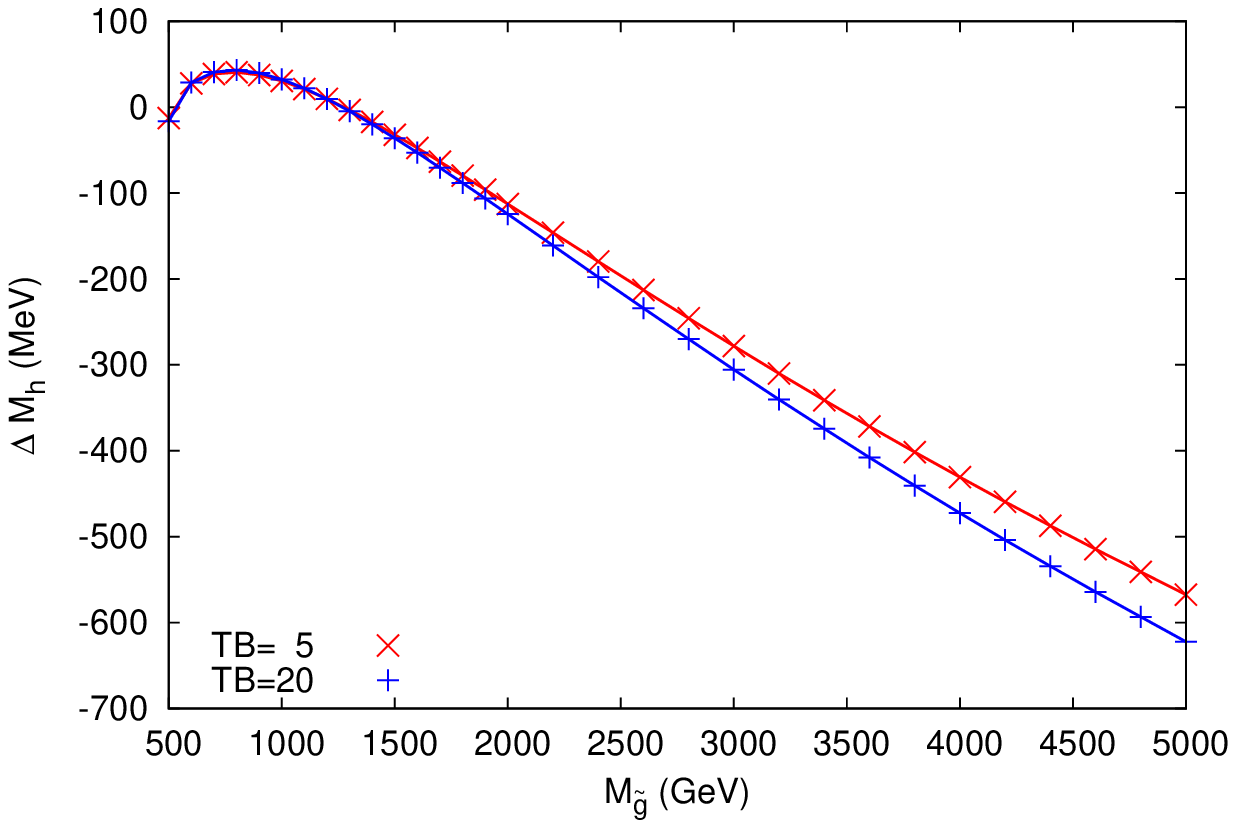}
\includegraphics[width=0.7\textwidth]{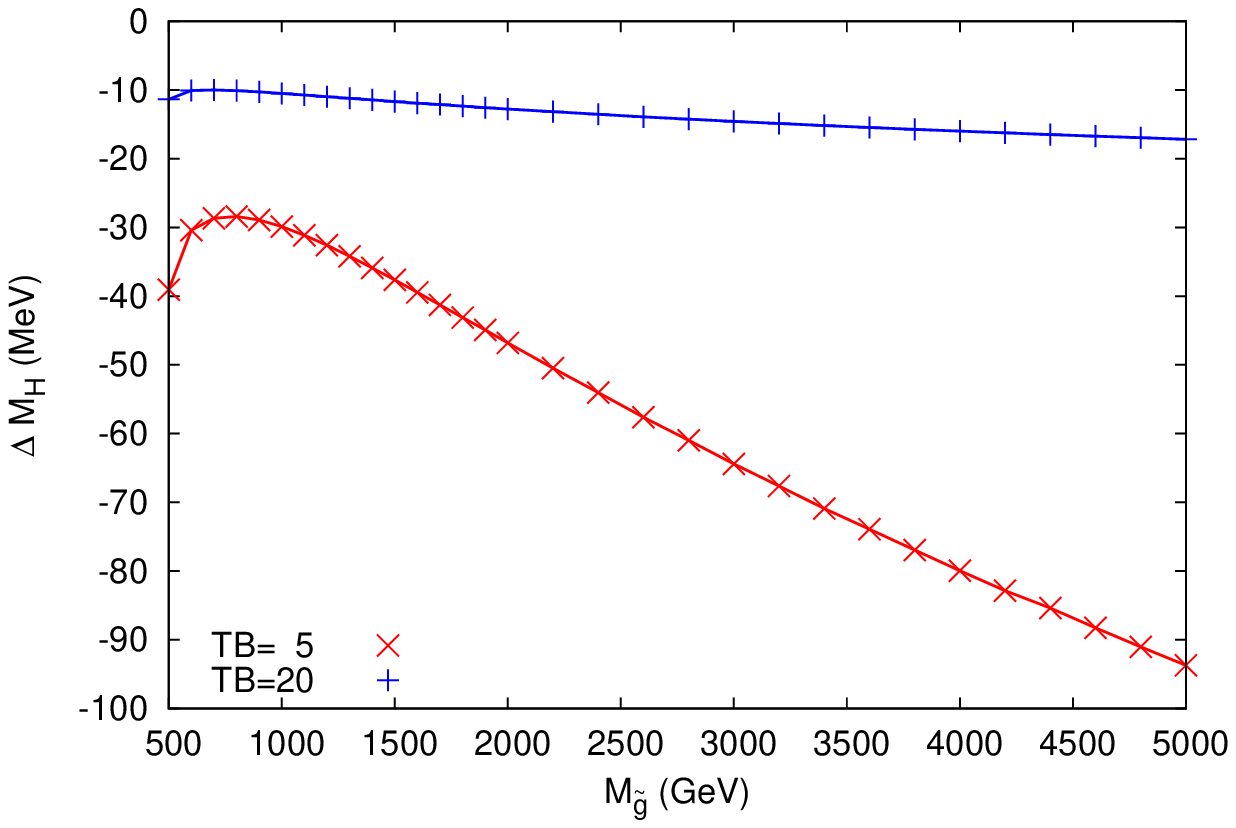}
\caption{Variation of the mass shifts $\Delta\Mh,\Delta\MH$ with the
  gluino mass, within scenario 2, for two different  values of
  $\tb=5,20$ and $\MA = 250 \gev$.
}
\label{fig:scen2variationmgluino}
\end{figure}

\medskip
Finally we analyze the dependence on $\Mgl$ in
\reffi{fig:scen2variationmgluino}. In the upper plot we show $\De\Mh$
for $\tb = 5$ and $\tb = 20$, where both values yield very similar
results. As in scenario~1, ``accidentally'' small values of $\De\Mh$ are
found around $\Mgl \sim 1600 \gev$. For larger gluino mass values the
shifts induced by the new momentum-dependent two-loop corrections exceed
$- 500 \mev$ and are thus larger than the current experimental
uncertainty. The results for $\De\MH$ are shown in the lower plot. While
they are roughly twice as large as in scenario~1, they do not exceed
$-100 \mev$.

\clearpage


\section{Conclusions}
\label{sec:conclusions}

We have presented results for the leading momentum-dependent
\order{\alt\als} contributions to the masses of neutral 
$\cp$-even Higgs-bosons in the MSSM. 
They are obtained by calculating
the corresponding contributions to the dressed Higgs-boson propagators 
obtained in the Feynman-diagrammatic approach using a mixed
on-shell/\DRbar\  renormalization scheme. 
In the Higgs sector a two-loop renormalization has to be carried out for 
the mass of the neutral Higgs bosons and the tadpole contributions. 
Furthermore, renormalization of the top/stop sector
at \order{\als} is needed entering at the two-loop level
via one-loop subrenormalization.
The diagrams were generated with  {\tt FeynArts} and reduced to a
set of basic integrals with the help of  {\tt FormCalc} and {\tt TwoCalc}.
The two-loop integrals which are analytically unknown have been 
calculated numerically with the program \sd.

\smallskip
We have analyzed numerically the effect of the new momentum-dependent
two-loop corrections on the predictions for
the $\cp$-even Higgs boson masses. This is
particularly important for the interpretation of  the scalar boson 
discovered at the LHC as the light $\cp$-even Higgs state of the MSSM.
While currently a
precision below the level of $\sim 500 \mev$ is reached, a reduction by
about an order of magnitude can be expected at the future $e^+e^-$
International Linear Collider (ILC). 

\smallskip
In our numerical analysis we found that the effects on the light
$\cp$-even Higgs boson mass, $\Mh$, depend strongly on the value of the
gluino mass, $\Mgl$. For values of $\Mgl \sim 1.5 \tev$ corrections to
$\Mh$ of about $ -50 \mev$ are found, at the level of the anticipated
future ILC accuracy. For very large gluino masses, $\Mgl \gtrsim 4 \tev$, 
on the other hand, substantially larger corrections are found, at the
level of the {\em current} experimental accuracy. Consequently, this
type of momentum dependent two-loop corrections should be taken into
account in precision analyses interpreting the discovered Higgs
boson in the MSSM.

\smallskip
For the heavy $\cp$-even Higgs boson mass, $\MH$, the effects are mostly
below current and future anticipated accuracies. Only close to
thresholds, e.g.\ around $p = 2\,\mste$, larger corrections
to $\MH$ around $\sim 1 \gev$ are found.

The new results of \order{\alt\als} 
have been implemented into the program \fh.
A detailed description of our calculation
will be presented in a forthcoming publication~\cite{Mh-secdec}.

\clearpage

\appendix
\section*{Appendix: Renormalization and counterterms}
\label{sec:appendix}

Renormalization and calculation of the renormalized self-energies 
is performed in the $(\Pe, \Pz)$ basis, which 
has the advantage that the mixing angle $\al$ does not appear and
expressions are in general simpler. 

\medskip
Field renormalization is perfomed by assigning one
renormalization constant for each doublet, 
\begin{align}
\label{rMSSM:HiggsDublettFeldren}
  \cHe \to (1 + \tfrac{1}{2} \dZ{\cHe} )\, \cHe, \quad
  \cHz \to (1 + \tfrac{1}{2} \dZ{\cHz} ) \cHz\,,
\end{align} 
which can be expanded to one- and two-loop order according to
\begin{align}
\label{rMSSM:Feldrenexpand}
 \dZ{\cHe} & =\, \dZo{\cHe} +  \dZt{\cHe} \, , \quad
  \dZ{\cHz}  \, =\,  \dZo{\cHz} + \dZt{\cHz}\,. 
\end{align}
The field renormalization constants appearing in~(\ref{rMSSM:renses_higgssector})
are then given by
\begin{align}
\dZ{\Pe}^{(i)} = \dZ{\cHe} ^{(i)}\,, \quad 
\dZ{\Pz}^{(i)} = \dZ{\cHz} ^{(i)}\, , \quad
\dZ{\Pe\Pz}^{(i)} = \edz (\dZ{\cHe}^{(i)} + \dZ{\cHz}^{(i)}  )   \, .
\end{align}

The mass counterterms $\delta m^{2 (i)}_{ab}$ in~(\ref{rMSSM:renses_higgssector})
are derived from the Higgs potential, including the tadpoles, by the following
parameter renormalization,
\begin{align}
\label{rMSSM:PhysParamRenorm}
  \MA^2 &\to \MA^2 + \dMAsqo + \dMAsqt,  
& \tade &\to \tade + \dtadeo + \dtadet, \\ 
  \MZ^2 &\to \MZ^2 + \dMZsqo + \dMZsqt,  
& \tadz &\to \tadz + \dtadzo + \dtadzt, \notag \\ 
     \tb & \to \tb \KL 1 + \de\tb^{(1)} + \de\tb^{(2)} \KR~. \notag
\end{align}
The parameters $\tade$ and $\tadz$ are 
the terms linear in $\Pe$ and $\Pz$ in the Higgs potential. 
The renormalization of the $Z$ mass $M_Z$ does not contribute to the   
$\mathcal{O}(\alpha_s \alpha_t)$ corrections we are pursuing here;
it is listed, however, for completeness.

\medskip
The basic renormalization constants for parameters and fields have to
be fixed by renormalization conditions according to a 
renormalization scheme. Here we choose the on-shell scheme for
the parameters and the \drbar\ scheme for field 
renormalization 
and give the expressions for the two-loop part.

\medskip
The tadpole coefficients are chosen to vanish at all orders; hence
their two-loop counterterms follow from  
\begin{align}
T_{1,2} ^{(2)}+ \de T_{1,2}^{(2)} = 0 \,  , \quad
\ie \quad
  \dtadet = -{\tadet}, \quad \dtadzt = -{\tadzt}\,,
\end{align}
where $\tadet$, $\tadzt$ are obtained from the two-loop tapole diagrams.
The two-loop renormalization constant of the $A$-boson mass reads 
\begin{align}
\label{rMSSM:mass_osdefinition}
  \quad \dMAsqt = \re \se{AA}^{(2)}(\MA^2) ,
\end{align}
in terms of the $A$-boson unrenormalized self-energy $\se{AA}$. 
The appearance of a non-zero momentum in the self-energy goes beyond
the \order{\alt\als} corrections evaluated in
\citeres{mhiggsletter,mhiggslong,mhiggsEP1}. 

\medskip
For the renormalization constants $\dZ{\cHe}$, 
$\dZ{\cHz}$ and $\de\tb$ 
several choices are possible, see the discussion in~\cite{tbren}. 
As shown there, the most convenient
choice is a \drbar\ renormalization of $\de\tb$, $\dZ{\cHe}$
and $\dZ{\cHz}$, which reads at the two-loop level
\begin{subequations}
\label{rMSSM:deltaZHiggsTB}
\begin{align}
  \dZ{\cHe}^{(2)} &= \dZ{\cHe}^{(2)\drbarm}
       \; = \; - \KKL \re \Sipt_{\Pe} \KKR^{\rm div}_{|p^2 = 0}\,, \\[.5em]
  \dZ{\cHz}^{(2)} &= \dZ{\cHz}^{(2)\drbarm} 
       \; = \; - \KKL \re \Sipt_{\Pz} \KKR^{\rm div}_{|p^2 = 0}\,, \\[.5em]
  \dtanbt &= \dtanb^{(2)\drbarm} 
       \; = \; \edz  \KL \dZ{\cHz}^{(2)} - \dZ{\cHe}^{(2)} \KR\,.
       \label{eq:tanbren}
\end{align}
\end{subequations}
The term in \refeq{eq:tanbren} is in general not the proper expression
beyond one-loop order even in the  \drbar\ scheme.
For our approximation, however, with only the top
Yukawa coupling at the two-loop level, 
it is the correct \drbar\ form~\cite{Sperling:2013eva}.

\medskip
The two-loop mass counterterms 
in the self-energies~(\ref{rMSSM:renses_higgssector})
are now expressed in terms of the parameter renormalization constants
determined above as follows,
\begin{subequations}
\label{masscounterterms}
\begin{align}
\label{dm1sq}
\dmesqt &= \,\dMZsqt \, \CQb + \dMAsqt \SQb \\
\nonumber &\quad - \dtadet \frac{e}{2 \MW \sw} \, \Cb (1 + \SQb) 
+ \dtadzt \frac{e}{2 \MW \sw} \, \CQb \Sbe \\
\nonumber &\quad 
+ 2\, \dtanbt \, \cos^2\!\beta \sin^2\!\beta \, (\MA^2 - \MZ^2)\,, \\
\label{dm12sq}
\dmezsqt &= - (\dMZsqt + \dMAsqt) \Sbe \Cb \\
\nonumber &\quad - \dtadet \frac{e}{2 \MW \sw} \, \SDb 
                                - \dtadzt \frac{e}{2 \MW \sw} \, \CDb \\
\nonumber &\quad 
                 -  \dtanbt\, \cos\beta \sin\beta \cos 2\beta \, 
                      (\MA^2 + \MZ^2)\,, \\
\label{dm2sq}
\dmzsqt &= \dMZsqt \SQb + \dMAsqt \CQb \\
\nonumber &\quad + \dtadet \frac{e}{2 \MW \sw} \, \SQb \cos\beta
                 - \dtadzt \frac{e}{2 \MW \sw} \, \Sbe (1 + \CQb) \\
\nonumber &\quad - 2\, \dtanbt \, \cos^2\!\beta \sin^2\!\beta \, (\MA^2 - \MZ^2)\,.
\end{align}
\end{subequations}
Note that the $Z$-mass counterterm is kept for completeness; it
does not contribute in the approximation of \order{\als\alt} considered
here.


\bigskip
\subsection*{Acknowledgements}

We thank S.~Pa\ss{}ehr for help with the interfaces to
{\tt TwoCalc} and {\tt FormCalc}, 
S.~di~Vita for numerical comparisons
and 
G.~Weiglein for helpful discussions. 
The work of S.H.\ was supported by the 
Spanish MICINN's Consolider-Ingenio 2010 Program under Grant MultiDark No.\ 
CSD2009-00064.


\end{document}